\documentclass[onecolumn,aps,nofootinbib,notitlepage]{revtex4-1}

\usepackage[margin=2cm]{geometry}
\usepackage{amsfonts}
\usepackage{amsmath}
\usepackage{amsthm}
\usepackage{amssymb}
\usepackage[svgnames]{xcolor}
\usepackage[hidelinks]{hyperref}
\hypersetup{
    colorlinks=true,
    linkcolor=NavyBlue,
    filecolor=NavyBlue,
    urlcolor=NavyBlue,
    citecolor=NavyBlue,
    }

\usepackage{graphics}
\usepackage{tikz}
\usepackage{graphicx}
\usetikzlibrary{positioning}
\usepackage{caption}
\usepackage{subcaption}
\usepackage{adjustbox}
\usepackage{gensymb}
\usepackage{breqn}
\usepackage{braket}
\usepackage{enumitem}
\usepackage{gensymb}
\usepackage[compat=1.0.0]{tikz-feynman}

\def\gev{\, {\rm GeV}}
\def\mev{\, {\rm MeV}}
\def\pb{\, {\rm pb}}
\def\cm{\, {\rm cm}}
\def\s{\, {\rm s}}

\def\sigmaSI{\sigma_{\rm SI}}

\def\be{\begin{equation}}
\def\ee{\end{equation}}
\def\bea{\begin{eqnarray}}
\def\eea{\end{eqnarray}}

\begin{document}

\begin{center}
\hfill  MI-TH-1921, UH511-1305-2019
\end{center}

\title{\large{\textbf{A sub-GeV dark matter model}}}
\author{\normalsize{ Bhaskar Dutta$^{\bf 1}$\footnote{dutta@physics.tamu.edu}, Sumit Ghosh$^{\bf 1}$\footnote{ghosh@tamu.edu}, Jason Kumar$^{\bf 2}$\footnote{jkumar@hawaii.edu}} \\
\vspace{1.0cm}
\normalsize\emph{$^{\bf 1}$Mitchell Institute for Fundamental Physics and Astronomy, Department of Physics  and Astronomy, Texas A$\&$M University,
College Station, Texas 77843, USA}\\
\normalsize\emph{$^{\bf 2}$Department of Physics, University of Hawaii, Honolulu, Hawaii 96822, USA }\\
\vspace{1.0cm}
}

\begin{abstract}
We propose an extension of the Standard Model gauge symmetry by the gauge group $U(1)_{T3R}$ in order to address the Yukawa coupling hierarchy between the third generation  and the first two generation fermions of the SM. We assume that only the right-handed fermions of the first two generations are charged under the $U(1)_{T3R}$. In addition to the new dark gauge boson, we have a dark scalar particle whose vacuum expectation value (vev) breaks the $U(1)_{T3R}$ symmetry down to $Z_2$ symmetry and also explains the hierarchy problem. A vev of $\cal O$(GeV) is required to explain the mass parameters of the light flavor sector naturally. The dark matter (DM) particle arising from the model naturally has mass in the ${\cal O}(1-100)~\mev$ range. The model satisfies all the current constraints. We discuss the various prospects of the direct detection of the dark matter. We get both elastic and inelastic spin independent DM-nucleon scattering. The dark matter obtains the correct thermal relic density by annihilation.\end{abstract}
\maketitle

\section{Introduction}

In recent years, the central motivation for dark matter direct detection, indirect detection, and collider search strategies has been the WIMP Miracle~\cite{Jungman:1995df}.  In this paradigm, the key observations are twofold.  First of all, the WIMP Miracle is the statement that a stable particle with a mass of ${\cal O}(100-1000)~\gev$, annihilating to Standard Model particles with an ${\cal O}(1)$ coupling, would have a thermal relic density which is consistent with cosmological dark matter.  But equally important is the fact that new physics is naturally expected at the ${\cal O}(100-1000)~\gev$ scale since that is the scale of electroweak symmetry breaking.  Essentially, that is the scale by which energies in the electroweak sector are measured, so if dark matter couples to the electroweak sector, then it is natural to find particles at the correct scale needed to invoke the WIMP Miracle.

But as experimental searches, so far, have failed to find conclusive evidence for WIMPs, there has been a new interest in models of dark matter with mass in the ${\cal O}(1-100)~\mev$ range.  These models can evade tight constraints from current direct detection, indirect detection,  collider searches and various  low energy experiments, but maybe detected with data from planned experiments. Moreover, a variety of new mechanisms which have recently been discussed through which a stable particle in the ${\cal O}(1-100)~\mev$ range could obtain the correct relic density (see, for example,~\cite{Hochberg:2014kqa,Kuflik:2015isi}).  But what has thus far been lacking is a natural reason to have a new particle at the MeV-scale, beyond the fact that this particle could be a dark matter candidate.  But there is, in fact, another natural scale associated with flavor physics of the two lightest generations.  The mass parameters of the charged Standard Model fermions in the two lightest generations all lie in the ${\cal O}(1-100)~\mev$ range, and if dark matter arises from new physics associated with a light flavor, then it will also naturally lie at that scale.  Our aim in this work is to present a concrete realization of this scenario, in which dark matter is part of the light flavor sector, with a mass naturally at the MeV-scale.

A natural way to implement this idea is by adding a new gauge group, $U(1)_{T3R}$, under which right-handed fermions of the first two generations have charge $\pm 2$.  In addition to the dark photon $A'$, there is a dark Higgs scalar field $\phi$ charged under $U(1)_{T3R}$, whose vacuum expectation value breaks $U(1)_{T3R}$ to a $Z_2$.  Finally, the dark matter, $\eta_1$, is the lightest fermion which is odd under this surviving parity.    In the low energy effective theory below the electroweak scale, the masses of the first generation SM fermions, as well as the dark sector particles, are all proportional to the vev of $\phi$.  To explain the light flavor sector mass parameters, a natural scale for this vev is ${\cal O}(\gev)$, implying that the dark sector particles, like the first- and second-generation mass parameters, should be sub-GeV scale \footnote{Note, the coupling of the dark sector to $U(1)_{T3R}$ was considered in~\cite{Dutta:2010va} for a related motivation, namely, to provide a single energy scale which sets the mass of SM fermions and dark sector particles.  In that work, the dark sector coupled to $b$, $c$ and $\tau$, and the mass scale of the dark sector was ${\cal O}(\gev)$, providing for a good asymmetric dark matter candidate.  We will see that coupling the dark sector to the light flavor sector instead naturally leads to a sub-GeV dark matter candidate.}.

This scenario yields a rich phenomenology.  There are tight constraints on this scenario emerging from various low energy measurements including constraints on first- and second-generation lepton dipole moments. But we will find models which can satisfy all current constraints, and for which the dark matter thermal relic density is sufficiently depleted by annihilation via the dark Higgs resonances or dark photon mediated process.  Direct detection signals are also striking. Because $U(1)_{T3R}$ is broken to a $Z_2$ which stabilizes the dark matter candidate, the dark sector naturally contains a Dirac fermion which is split into two Majorana fermions, one or both of which are DM candidates.  Spin-independent (SI) DM-nucleon scattering can thus proceed by two methods, elastic scattering mediated by the dark Higgs, and inelastic scattering mediated by the dark photon.

The plan for this paper is as follows: In Sec.~\ref{Model}, we discuss the model building and all the necessary interaction terms. We discuss various constraints relevant to our model in Sec.~\ref{Constraints}. In Sec.~\ref{Specific Model} we choose two specific models for consideration, based on the constraints. In Sec.~\ref{Direct Detection} we discuss the direct detection prospects of our model. Sec.~\ref{Relic density} is about the relic density calculation. We conclude in Sec.~\ref{conclusion}. We provide additional details in the appendices. Appendix~\ref{appendix1} provides information about the nuclear form factor and dark matter velocity distribution. In Appendix~\ref{appendix2} we provide some details about the relic density calculation.

\section{Model} \label{Model}

The low energy gauge symmetry of our model is $SU(3)_C$$\times$$SU(2)_L$$\times$$U(1)_Y$$\times$$U(1)_{T_{3R}}$. We will assume that the new gauge group $U(1)_{T_{3R}}$ is not connected to electric charge, defined as $Q$$=$$T_{3L}$$+$$Y$. But we note that one can also consider this scenario in the context of left-right models, in which case the hypercharge $Y$ is determined by the charge under $U(1)_{B-L}$ and $U(1)_{T3R}$. We assume that only the  right-handed Standard Model (SM) fermions (including the right-handed neutrinos) are charged under the $U(1)_{T_{3R}}$ gauge group.   We assume that no other SM fields are charged under $U(1)_{T_{3R}}$, and all SM fields have their usual charges under the SM gauge groups. In addition to the right-handed neutrinos, there will be three other new matter fields, a scalar $\phi$, and a left and right-handed fermion pair $\eta_L$ and $\eta_R$. These new matter fields are SM singlets and only charged under $U(1)_{T_{3R}}$. There is also a new gauge boson, the dark photon $A'$. All fields with non-trivial charges under $U(1)_{T3R}$ are listed in Table~\ref{tab:charges}. Since we describe both left-handed and right-handed Weyl fields, for consistency and clarity we list the charges of the left-handed component of the Weyl spinor.  Note that, as expected, all gauge anomalies cancel if the SM fields charged under $U(1)_{T3R}$ consist of a full right-handed generation, including an up-type quark, down-type quark, charged lepton, and a neutrino.  Thus, this model is anomaly-free if either one or two generations couples to $U(1)_{T3R}$.

\begin{table}[h]															 \centering															\begin{tabular}{ |c|c|c|c|c||c|c|c| }											\hline																			\hline																			field&$q_R^u$&$q_R^d$& $\ell_R$&$\nu_R$&$\eta_L$&$\eta_R$&$\phi$ \\ \hline		$q_{T3R}$&-2&2&2&-2&1&-1&-2\\\hline\hline								\end{tabular}															\caption{The charges of fields which transform under $U(1)_{T3R}$. For fermions, the charges are given for the left-handed component of each Weyl spinor. The anomalies cancel by construction.} \label{tab:charges}										\end{table}

The Yukawa interactions for the new fields can be written in terms of the Lagrangian,
\begin{dmath} \label{intlag} \mathcal{L}_{Yuk}=-\frac{\lambda_u}{\Lambda} \tilde{H} \phi^* \bar{Q}_L  q_R^u  -  \frac{\lambda_d}{\Lambda} H \phi \bar{Q}_L q_R^d - \frac{\lambda_\nu}{\Lambda} \tilde{H} \phi^*\bar{L}_L  \nu_R  - \frac{\lambda_\ell}{\Lambda} H \phi \bar{L}_L \ell_R - m_D \bar{\eta}_R  \eta_L - \frac{1}{2}\lambda_L \phi \bar{\eta}^c_L \eta_L - \frac{1}{2}\lambda_R \phi^* \bar{\eta}^c_R \eta_R-\mu_\phi^2 \phi^* \phi - \lambda_\phi (\phi^* \phi)^2 +H.c. , \end{dmath}
where $Q_L$ and $L_L$ are the left-handed SM quark and lepton doublet, respectively. $H$ is the SM Higgs doublet and $\tilde{H}$$=$$i\tau_2H^*$. 

The potential terms in Eq.~(\ref{intlag}) will cause $\phi$ to get a vacuum expectation value(vev), $V$$=$$(-\mu_\phi^2/2 \lambda_\phi)^{1/2}$ and will yield one physical real scalar field $\phi^\prime$ with mass $m_{\phi^\prime}$$=$$2\lambda_\phi^{1/2} V$. The vev will break the gauge group $U(1)_{T{3R}}$ down to a $Z_2$ symmetry group, under which $\phi^\prime$ and all of the SM fields are even. Only $\eta_{L,R}$ will be odd under this parity.

The first four terms of Eq.~(\ref{intlag}) will give the mass terms for the up type quark, the down type quark, and the charged lepton and the tree level Dirac mass term of the neutrino. They also give the interaction terms of the quarks and leptons with the physical scalar $\phi^\prime$. The $\eta$ field will get both Dirac and Majorana masses. For simplicity, we assume $\lambda_L = \lambda_R \equiv \lambda_M$.  This is the maximal mixing case. The Majorana masses for the left-handed and the right-handed fields are equal, with $m_M=\lambda_L V= \lambda_RV=(\lambda_M V)$. We consider the Dirac mass, $m_D$ to be very small compared to the Majorana mass, $m_M$. We get two physical Majorana fields  which are,
\bea \eta_1 &=& -\frac{1}{\sqrt{2}}\left( \begin{array}{c} \eta_L-\eta^c_R \\ -\eta^c_L+\eta_R \end{array} \right) ,                                    \nonumber\\                                                                   \eta_2 &=& \frac{1}{\sqrt{2}}\left( \begin{array}{c} \eta_L+\eta^c_R \\ \eta^c_L+\eta_R \end{array} \right) ,    													\eea
with the corresponding real and positive masses  $m_1 = m_M-m_D $ and $m_2=m_M+m_D$ respectively. The mass splitting between them is $\delta = 2m_D$, which is very small. The small Dirac mass term ensures that the couplings of $\phi^\prime$ to $\eta_{1,2}$ are proportional to the mass $m_{1,2}$. The two Majorana physical fields $\eta_{1,2}$ are the dark matter fields in our model. The lightest of them will be absolutely stable.

We can then rewrite the Eq.~(\ref{intlag}) in terms of the  mass of the  physical quarks and leptons and the dark Higgs vev $V$.
\bea \mathcal{L}_{Yuk} &=& -m_u\bar{q}^u_L q^u_R  -m_d\bar{q}^d_L q^d_R  -m_{\nu D}\bar{\nu}_L \nu_R -m_\ell \bar{\ell}_L \ell_R -\frac{1}{2} m_1 \bar{\eta}_1  {\eta}_1 -\frac{1}{2}m_2 \bar{\eta}_2  {\eta}_2								\nonumber\\																		&\,& -\frac{m_u }{V}\bar{q}^u_L q^u_R \phi^\prime -\frac{m_d  }{V}\bar{q}^d_L q^d_R \phi^\prime -\frac{m_{\nu D}  }{V}\bar{\nu}_L \nu_R \phi^\prime	-\frac{m_\ell }{V}\bar{\ell}_L \ell_R \phi^\prime -\frac{1}{2}\frac{m_1}{V} \bar{\eta}_1  {\eta}_1 \phi^\prime-\frac{1}{2}\frac{m_2}{V} \bar{\eta}_2  {\eta}_2 \phi^\prime + ....     \eea

To explore the new gauge sector of our model we first define the covariant derivative as,
\begin{equation} D_{\mu} = {\partial}_{\mu} +i\frac{g}{2}{\tau}_a W_{\mu a}+ig^{\prime} Y B_{\mu} +i\frac{g_{T_{3R}}}{2}Q_{T_{3R}} A'_\mu , \end{equation} where $g$, $g^{\prime}$ and $g_{T_{3R}}$ are the coupling constant corresponding to the $SU(2)_L$, $U(1)_Y$ and $U(1)_{T_{3R}}$ groups respectively. $W_{\mu }$, $ B_{\mu}$ and $A'_\mu$ are the gauge bosons of the $SU(2)_L$, $U(1)_Y$ and $U(1)_{T_{3R}}$ groups respectively.

$|D_{\mu}H|^2$ gives the masses of the SM gauge bosons $W^{\pm}$ and $Z$, while $|D_{\mu}\phi|^2$ gives the mass of the dark photon $A^\prime$, yielding $m_{A^\prime}^2=2g_{T_{3R}}^2 V^2$. The trilinear interactions involving the gauge boson $A^\prime$ are then given by,
\bea  \mathcal{L}_{gauge} \label{lgauge} =  \frac{1}{4}g_{T_{3R}}A^\prime_\mu(\bar{\eta}_1\gamma^\mu\eta_2-\bar{\eta}_2\gamma^\mu\eta_1)  +  \frac{m_{A'}^2}{V} \phi' A'_\mu A'{}^{\mu} +i g_{T3R} A'_\mu \left(\phi' \partial^\mu \phi'{}^{*} - \phi'{}^{*} \partial^\mu \phi \right) -\frac{1}{2} g_{T_{3R}} j^\mu_{A^\prime}{A^\prime}_\mu ,     \eea
where the SM  interaction current is defined as, $j^\mu_{A^\prime}=\sum\limits_f Q_{T_{3R}}^f \bar{f}\gamma^\mu f$. Only off-diagonal vector interaction terms exist for the Majorana dark matter fields.

\begin{figure}[h]														\begin{subfigure}[b]{0.48\textwidth}							\includegraphics[width=.8\linewidth,height=3cm]{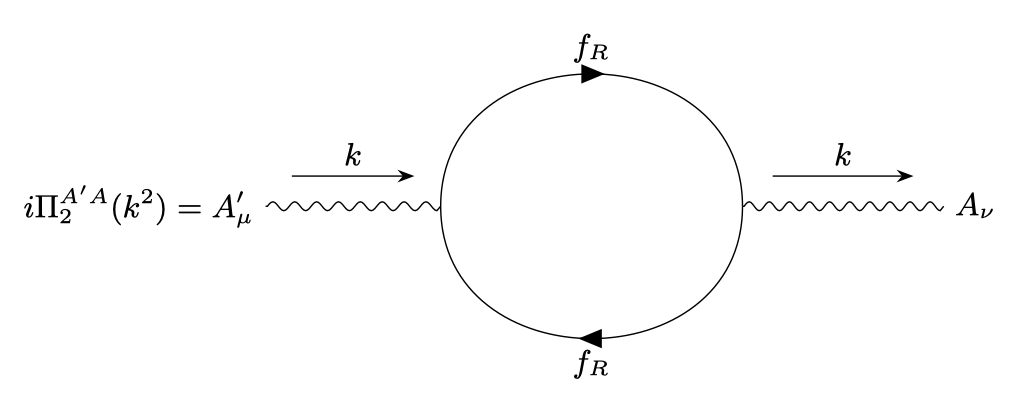}\caption{\label{fig:loop1}}											\end{subfigure}															\hspace{0.2cm}														\begin{subfigure}[b]{0.48\textwidth}\includegraphics[width=0.8\linewidth,height=3cm]{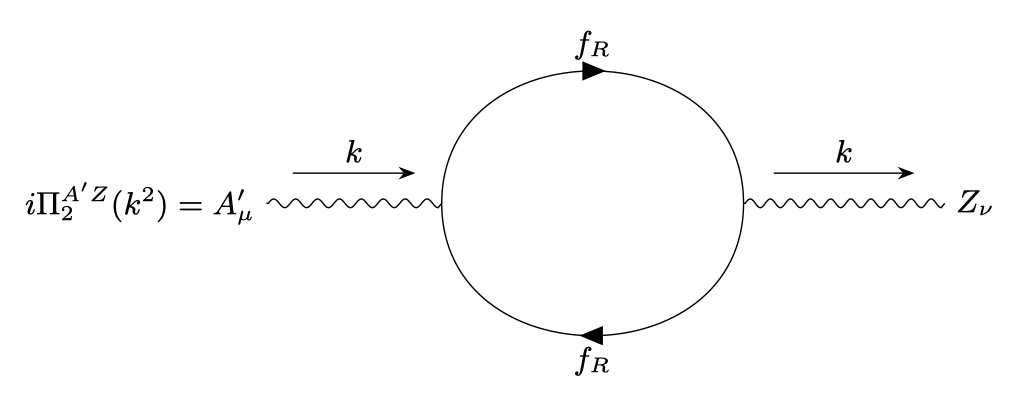}\caption{\label{fig:loop2}}												\end{subfigure}													\caption{\label{fig:loop} The one loop diagrams  which give the mixing induced coupling between the SM fields to $A^\prime$. Here, $f_R= \ell_R, q^u_R, q^d_R$.}\end{figure}

The dark photon, $A^\prime$ will mix with the photon and the $Z$ boson  due to the diagrams in which $q^{u, d}_R$ and $\ell_R$ run in the loop (Fig.~\ref{fig:loop}) \footnote{We have used the package TikZ-Feynman~\cite{Ellis:2016jkw} to draw the diagram.}. We assume that there is no tree-level kinetic mixing between them. As a result, even the SM fields which are uncharged under $U(1)_{T_{3R}}$ will get a mixing induced coupling to $A^\prime$. The kinetic mixing becomes smaller as the mass of $A^\prime$ decreases.

For simplicity, we will assume that $U(1)_{T3R}$ couples only to one charged lepton, one up-type quark,  one down-type quark, and one neutrino (all right-handed). We further assume that the charged lepton and down-type quark states are mass eigenstates, while the up-type quark state is a linear combination of all up-type mass eigenstates.  This coupling structure is technically natural, as it yields an extra $U(1)^2$ flavor symmetry arising from rotations of the charged lepton  and down-type quark wave functions by an independent phase~\cite{Batell:2017kty}.  Note, there is no additional restored symmetry if we take the up-type quark state to also be a mass eigenstate, due to the CKM matrix. Note that, beyond the dark matter candidates $\eta_{1,2}$, we have introduced three other new particles: $A'$, $\phi'$ and $\nu_R$. Since $\nu_R$ is charged under no unbroken symmetries, it will generically mix with the left-handed neutrinos, yielding several Majorana neutrino mass eigenstates.  We will assume that $\nu_R$ is dominantly composed of a sterile neutrino mass eigenstate $\nu_s$, with some mixing with the active neutrino mass eigenstates (collectively denoted by $\nu_A$). The mass of the lightest neutrino eigenstate can be determined by the seesaw mechanism. If there is no Majorana mass for the left-handed neutrinos, then the masses of the eigenstates determine the neutrino mixing angle. However, the mixing angle is not fixed if the left-handed neutrino also has a Majorana mass term. None of the new particles are stabilized by any symmetry (assuming, as we do, that $\nu_s$ is not the lightest fermion), and thus they should all be able to decay to Standard Model particles.  The main decay channels for these particles are
\begin{itemize}																	\item{$\phi'$: $\phi' \rightarrow \bar \ell \ell, \nu_s \nu_A, \pi \pi, A' A'$ dominate, if kinematically allowed. If those tree-level decays are not allowed, then $\phi' \rightarrow \gamma \gamma$ (mediated by a one-loop diagram) will dominate.}																	\item{$A'$: $A' \rightarrow \bar \ell \ell, \nu_s \nu_s, \pi \pi, \phi' \phi'$ dominate, if kinematically allowed. If they are not kinematically allowed, $A' \rightarrow \nu_A \nu_A$ will dominate.}										\item{$\nu_s$: $\nu_s \rightarrow \nu_A \gamma \gamma$ (mediated by an off-shell $\phi'$) will dominate.} 													\end{itemize}

\section{Constraints} \label{Constraints}

In this section, we discuss various constraints for this model.  Constraints on the coupling of a dark photon to Standard Model particles are discussed in~\cite{Bauer:2018onh}, while constraints on the coupling of a dark Higgs to Standard Model particles are discussed in~\cite{Batell:2017kty}. From Eq.~(\ref{intlag}) we see that $H$, $q^{u,d}$, $\ell$ and $\nu$ couple to $\phi$.  In addition, $q^{u,d}$, $\ell$ and $\nu$ couple to $A'$, the gauge boson of $U(1)_{T3R}$, with coupling $g_{T3R}$.  Since the couplings of all new particles to $\nu_A \nu_A$ are suppressed by a mixing angle, we will find suppressed constraints from Borexino~\cite{Kouda:2016ifp, Bellini:2011rx, Harnik:2012ni}, Texono~\cite{Deniz:2009mu}, Coherent~\cite{Akimov:2015nza, Akimov:2017ade}, Charm II~\cite{Vilain:1994qy, Vilain:1993kd}, NuTeV~\cite{Adams:1998yf}, CCFR~\cite{Mishra:1991bv}, etc. We thus have to consider constraints on the following processes.

\begin{enumerate}[label=(\roman*)]

\item{{\it Corrections to the lepton magnetic dipole moments.} The lepton magnetic dipole moments will receive corrections from one-loop diagrams involving either the dark photon or the dark Higgs. The correction to $a_\ell = (g_\ell -2)/2$  due to one-loop diagrams involving $A'$ and $\phi'$ is given by~\cite{Leveille:1977rc} \bea \label{delta} \delta a_\ell &=& \frac{m_\ell^4}{8\pi^2V^2}  \int_0^1  \frac{(1-x)^2(1+x)}{(1-x)^2m_\ell^2+xm_{\phi^\prime}^2}  dx + \frac{m_\ell^2 }{32\pi^2V^2} \int_0^1  \frac{2x(1-x)(x-2)m_{A^\prime}^2-2x^3m_\ell^2}{x^2m_\ell^2+(1-x)m_{A^\prime}^2} dx .  \eea But one must stress that there can be additional contributions to the lepton dipole moments from heavy new physics unrelated to the light flavor sector.  As such, the dipole moment constraints are not true constraints on the model, but rather measure the level of fine-tuning in the cancellation between corrections from the light flavor sector those from heavy new physics.}

\item {{\it Constraints from $e^+ e^-$ colliders}: BaBar~\cite{Aubert:2009cp, Lees:2014xha} and/or  Belle~\cite{Abe:2010gxa, Inguglia:2016acz}  constrain dark photon and dark Higgs couplings by searching for the process $e^+ e^- \rightarrow A^\prime,\,\phi' \rightarrow \mu^+\mu^-, e^+ e^-, \nu\nu$, and  $e^+ e^- \rightarrow \mu^+\mu^-+A^\prime,\,\phi' \rightarrow 4 \mu$, while KLOE~\cite{Archilli:2011zc, Babusci:2012cr, Anastasi:2016ktq, Anastasi:2015qla} can provide similar constraints with the process $e^+ e^- \rightarrow A^\prime,\,\phi' \rightarrow \mu^+\mu^-$. }

\item{{\it Anomalous $\pi^0$ decay}: Proton beam-dump experiments, such as LSND~\cite{Athanassopoulos:1997er} and NA 48/2~\cite{Batley:2015lha}, provide constraints on neutral pion production, followed by the decay  $\pi^0 \rightarrow \gamma (A',\phi') \rightarrow \gamma e^+ e^-$.}

\item{ {\it Invisible $A',\phi'$ decay:} NA64~\cite{Banerjee:2016tad, Gninenko:2018tlp, Gninenko:2014pea} constrains models in which $A^\prime$, $\phi'$-strahlung arising from the electron, followed by invisible decay ($A',\phi\rightarrow  \nu\nu$ or $\phi' \rightarrow A' A'$), yields missing energy.}

\item{{\it $A',\phi' \rightarrow e^+ e^-$}: Electron and proton beam dump experiments, including E137~\cite{Riordan:1987aw,Bjorken:1988as,Bjorken:2009mm,Andreas:2012mt}, E774~\cite{Bross:1989mp}, Orsay~\cite{Davier:1989wz}, LSND~\cite{Athanassopoulos:1997er} etc., can search for processes in which $A'$ or $\phi'$ is produced by bremsstrahlung, and later decays via $A', \phi' \rightarrow e^+ e^-$.}

\item{{\it Constraints arising from atomic parity violation experiment}~\cite{Diener:2011jt}. The nuclear transition $6S_{1/2}$-$7 S_{1/2}$ in $^{133}Cs$, which is allowed by the parity violation, has been measured by multiple collaborations to great precision.}

\item{ {\it Constraints arising from the cooling of white dwarfs and supernovae}~\cite{Chang:2018rso, Dreiner:2013tja, Harnik:2012ni}. The light mediators ($\phi'$  and $A'$) may be produced in the core of a supernova or white dwarf, and contribute to its energy loss from the coupling to neutrinos and dark matter.}

\item{{\it Globular cluster and solar capture constraints}~\cite{Harnik:2012ni}:  Solar energy loss and cooling of stars in globular clusters can occur due to the production of $\phi^\prime$ and $A'$ which decay into dark matter or neutrinos.}

\item{{\it Big-Bang Nucleosynthesis (BBN)}: Limits from  BBN ~\cite{Mangano:2011ar} on the effective number of new relativistic degrees of freedom (beyond three neutrinos), $\Delta N_{eff}\leq 0.2-0.6$, provide constraints on the parameter space when we have  particles with mass $\leq$ MeV.}

\item{{\it Fifth force}~\cite{Bordag:2001qi,Adelberger:2006dh, Adelberger:2009zz}: A new long-range force is constrained by precision tests of the gravitational, Casimir, and van der Waals forces.  These experiments are sensitive to $U(1)_{T3R}$ forces since they probe interactions between electrically neutral objects.}

\end{enumerate}

The LHC constraints on the Higgs decay process $H \rightarrow \bar f f \phi'$ may need to be considered.  Although the decay of the Higgs to first- or second-generation fermions is suppressed by the factor $m_f^2 / v^2$ (where $v \sim 246 ~ \gev$ is the Higgs vev), the decay to $\bar f f \phi'$ is also enhanced by an extra factor of $v^2 / V^2$. But  this factor is also compensated by the  additional 3-body phase space factor $\sim 1/(16 \pi^2)$; for $V\sim 10~\gev$, the Higgs decay process $H \rightarrow \bar f f \phi'$ is still negligible. This Model satisfies all constraints from precision electroweak data~\cite{Tanabashi:2018oca}.

\section{Specific Model} \label{Specific Model}

In this section, we consider specific models, chosen for simplicity,  to naturally get MeV-scale dark matter and satisfy flavor constraints. The mass of a first - or second-generation SM fermion which couples to the dark sector obeys the relation $m_f = \lambda_{\phi' \bar f f} V$, where $\lambda_{\phi' \bar f f}$ is the coupling of the SM fermion to the physical dark Higgs. As a result of the light flavor sector we have introduced, the mass of the SM fermion is set by the energy scale $V$; assuming there is no additional relevant flavor physics to further suppress the SM fermion mass, we would expect  $m_f \lesssim V$, implying  $V \sim {\cal O}(1-10)~\gev$. Since $ m_{\phi^\prime}^2 = 4 \lambda_\phi V^2 $ and $m_{A'}^2 = 2 g_{T3R}^2 V^2$, we will also get $ m_\phi^\prime, m_{A'} \lesssim {\cal O}$(GeV).

Under the scenario we consider, the down-type quark and charged lepton states which couple to $U(1)_{T3R}$ are mass eigenstates, while the up-type quark and neutrino states need not be.  For simplicity, we focus on the case in which the down-type quark state charged under $U(1)_{T3R}$ is the $d$ mass eigenstate (though we will discuss other possibilities). Although the up-type quark state charged under $U(1)_{T3R}$ can be a linear combination of $u$ and $c$ mass eigenstates, we assume for simplicity that the contribution from the $c$ mass eigenstate is negligible (such a contribution would have negligible effect on direct detection sensitivity, but might allow rare charm meson decays which are constrained by data).  The states which couple to the dark sector are thus $u$, $d$, a single right-handed neutrino, and either $\mu$ or $e$.

We will take $V \sim 10~$GeV, and will assume that the neutrino mixing angle is small. We then have two scenarios:
\begin{itemize}	
															
\item{$\ell=\mu$: In this case, the $\phi' \mu \mu$ coupling is $\sim 0.01$.
A relevant constraint for this case, if $m_{\phi^\prime}\ge 5$ GeV, arises from the analysis of $4\mu$ final states by CMS~\cite{Sirunyan:2018nnz}. For the $\phi' \mu \mu$ coupling equal to $0.01$, the constraint can be evaded if the branching ratio, $Br(\phi^\prime \rightarrow \mu \mu ) < 1/4$, which is the case for our model since $\phi^\prime$ mostly decays to $A' _\mu A'_\mu$. The branching ratio $Br(\phi^\prime \rightarrow \mu \mu )$ is $\mathcal{O}(10^{-4})$.  This constraint is not relevant for $m_{\phi^\prime} < 5$ GeV.
For $2m_\mu \leq m_{\phi'} \lesssim 5~\gev$, our scenario would be ruled out by constraints from BaBar on the process $e^+ e^- \rightarrow \mu^+ \mu^- \phi' (\phi' \rightarrow \mu^+ \mu^-)$~\cite{Batell:2017kty}. The mass range $1~\mev \lesssim m_{\phi^\prime} \lesssim 50~\mev$ is ruled out by constraints from E137 on the production of long-lived particles which decay to $\gamma \gamma$ (in this case, the production of $\phi'$ and its subsequent decay are mediated by the operator $\phi' F^{\mu \nu} F_{\mu \nu}$, which is generated at one-loop~\cite{Batell:2017kty}). For $m_{\phi'} \lesssim 1~\mev$, $\phi'$ will decay dominantly to prompt photons if $\nu_s$ is taken sufficiently heavy; our scenario is thus unconstrained by bounds on the cooling of astrophysical bodies~\cite{Harnik:2012ni}. But for $m_{\phi'} \lesssim 10^{-9}$ GeV, our scenario is ruled out by constraints on a fifth force~\cite{Harnik:2012ni}.
    If  $\Delta N_{eff} \sim 0.2-0.6$ then both BBN and CMB constraints on extra light degrees of freedom are satisfied~\cite{Knapen:2017xzo, Cyburt:2015mya, Aghanim:2018eyx, Riess:2019cxk}.
    Since $\phi' \mu \mu $ coupling is large $(\sim 0.01)$, this constraint effectively rules out models of very light $\phi'$. 
From Eq.~\ref{delta}, we get that the $A^\prime$ correction to  $g_\mu-2$ is always large (given $V = $ 10 GeV) and negative, while the $\phi^\prime$ correction is always positive. If the $\phi^\prime$ is light enough, then its correction to $g_\mu-2$ is also large and can cancel the large, negative contribution from the gauge boson correction. But if the $\phi^\prime$ is too heavy, then the correction is too small to cancel the gauge boson correction and we need fine-tuning from new physics instead.

In the range $m_{A^\prime}\sim 0.004 -0.2 ~\gev$ our scenario  satisfies all the constraints~\cite{Bauer:2018onh}. The parameter space $m_{A^\prime} >$ 0.2 GeV  is ruled out by BaBar~\cite{Bauer:2018onh}. Since $A'$ couples to $\nu_R$, constraints from Borexino, COHERENT, CCFR and Charm-II are suppressed when the neutrino mixing angle is taken to be small. This parameter space is allowed by  the meson decay processes~\cite{laha, Raggi:2015yfk, deNiverville:2015mwa}. Constraints from  white dwarf cooling are also negligible if $m_\eta, m_{\nu_s} \gtrsim 0.1~\mev$~ \cite{Harnik:2012ni}, in which case the only available cooling process involves the coupling $ ee\nu_A \nu_A$, which is two-loop suppressed. The range  $m_{A^\prime}\sim 10^{-8}-10^{-3} ~\gev$ is ruled out by globular cluster, solar  and supernova cooling  constraints~\cite{Harnik:2012ni}, but these constraints can be relaxed due to chameleon effects~\cite{nelson, Feldman:2006wg, Nelson:2008tn}. The $m_{A'}\leq 0.004$ GeV is ruled out by E774~\cite{Bauer:2018onh}. The $m_{A'}\leq 10^{-9}$ GeV is ruled out by the fifth force constraints~\cite{Harnik:2012ni}.
As with the $\phi'$, the region of parameter space with very light $A'$ is also constrained by BBN and CMB bounds on $\Delta N_{eff}$.
This  constraint rules out the region of parameter space with  $5~\mev \gtrsim m_{A'} \gtrsim 0.1$ keV~\cite{Escudero:2019gzq}.
But, unlike the $\phi'$,  the coupling of $A'$ to matter scales with $m_{A'}$; we find that $A'$ would not be in equilibrium with the SM particle at early times
for $m_{A'}\lesssim 0.1$ keV, given our choice of $V=10$ GeV.
Moreover, since $A'$ is not directly connected to active neutrinos, it is not regenerated by them in inverse decay. However, this region is already ruled out by globular cluster, solar and supernovae cooling and E774 data. We choose two Benchmark Points (BP) which satisfy all the constraints and are later shown to have the correct relic density.  Both of our BP are at larger values of $m_{A'}$.

We show the allowed regions of $m_{\phi^\prime}$-$m_{A^\prime}$ parameter space for the $\ell = \mu$ case in Fig.~\ref{fig:muonallowed}. }

\begin{figure}[htb]													\begin{tikzpicture}																\node (img1)  {\includegraphics[height=8cm,width=14.2cm]{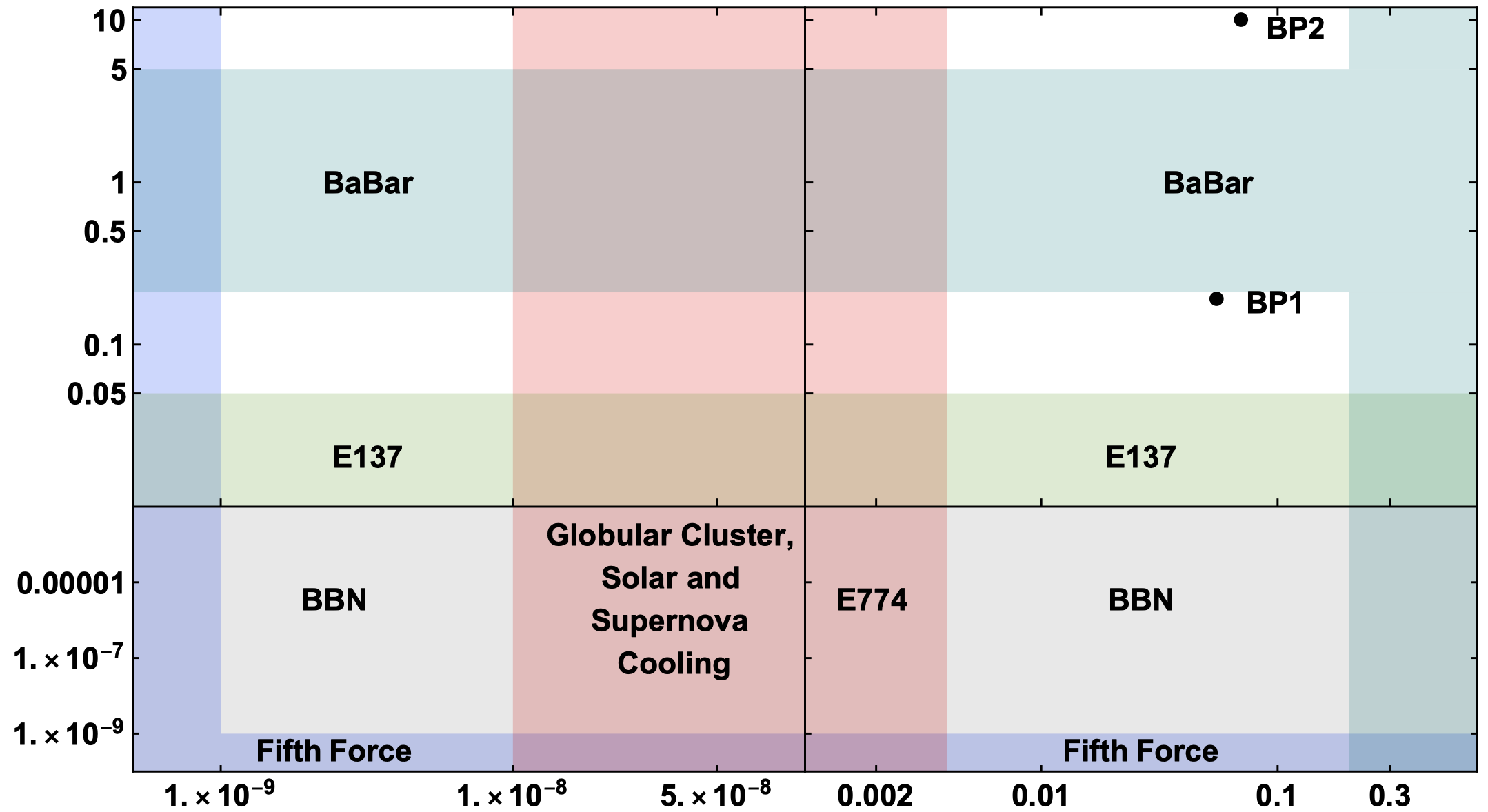}};		\node[below=of img1, node distance=0cm, yshift=.7cm,font=\color{black}] {\textbf{\large{$ m_{A^\prime}$} (GeV)}};								\node[left=of img1, node distance=0cm, rotate=90, anchor=center,yshift=-0.8cm,font=\color{black}] {\textbf{\large{$m_{\phi^\prime}$} (GeV)}};															\end{tikzpicture}											\caption{\label{fig:muonallowed} Allowed regions of $m_{\phi^\prime}$-$m_{A^\prime}$ parameter space for the case $\ell = \mu$ are shown as white region.  We assume that  $m_\eta , m_{\nu_s} \gtrsim 1~\mev$. We show two BP , BP1 : $m_{\phi^\prime}=$ 200 MeV and $m_{A^\prime}=$ 55 MeV and BP2 : $m_{\phi^\prime}=$ 10 GeV and $m_{A^\prime}=$ 70 MeV }				\end{figure}

\item{$\ell=e$: The  $\phi^\prime e e$ coupling for this scenario is $5\times 10^{-5}$, and any mass in the range $m_{\phi'} > 20~\mev$ is consistent with all constraints~\cite{Bauer:2018onh, Harnik:2012ni}. For $1~\mev \lesssim m_{\phi'} \lesssim 20~\mev$, this scenario can be constrained by searches at Orsay for $\phi'$ production, followed by the decay $\phi' \rightarrow e^+ e^-$~\cite{Bauer:2018onh} (though this bound is developed for the case of a coupling to the vector mediator, constraints for the case of a scalar mediator are comparable). For $10^{-6}$~GeV $\lesssim m_{\phi'} \lesssim10^{-4}$~GeV, this scenario is constrained by supernova cooling bounds~\cite{Harnik:2012ni}. However, if $\phi'$ decays into two photons promptly (which occurs if $\nu_s$ is heavier than $m_{\phi'}/2$) then the supernova constraint does not exist. The $m_{\phi'}\leq 10^{-7}$ GeV region is ruled out by the fifth force constraints~\cite{Harnik:2012ni}. The region $m_\phi^\prime \le 1 \mev$ is ruled out by the constraint $\Delta N_{eff}\sim 0.2- 0.6$~\cite{Knapen:2017xzo, Cyburt:2015mya, Aghanim:2018eyx, Riess:2019cxk}.

Constraints from atomic parity violation experiments exclude models with $m_{A'} \gtrsim 10~\mev$~\cite{Diener:2011jt, Abdullah:2018ykz} (taking the energy scale of the APV experiments to be 30 MeV), while the region  $1~\mev \leq m_{A^\prime}\leq10$ MeV is mostly ruled out by fixed target experiments, e.g., E774, E141, E137 etc~\cite{Bauer:2018onh}. The mixing between the active and sterile neutrinos can be assumed to be small.  There exist constraints from globular cluster and solar cooling for  $A^\prime$ masses in the ranges $10^{-4}$ GeV to $10^{-6}$ GeV and $10^{-7}$ GeV to $10^{-8}$ GeV, respectively~\cite{Harnik:2012ni}.  However, these constraints can be relaxed due to chameleon effects~\cite{nelson, Feldman:2006wg, Nelson:2008tn}. For $m_{A'} \lesssim 10^{-9}$ GeV, our scenario is again constrained by bounds on a fifth force~\cite{Harnik:2012ni}. The constraint   $\Delta N_{eff}\sim 0.2- 0.6$~\cite{Knapen:2017xzo, Cyburt:2015mya, Aghanim:2018eyx, Riess:2019cxk}  rules out the parameter space for $m_{A'}\leq 5$ MeV down  to  $m_{A'}\sim 0.01$ keV~\cite{Escudero:2019gzq}. $A'$ was not in equilibrium with the SM particles at early times for $m_{A'}\sim 0.01$ keV for our choice of $V$=10 GeV. $A'$ does not get regenerated by inverse decay. We show this region with a very light shaded region in the $m_{\phi^\prime}$-$m_{A^\prime}$ parameter space. Similarly, $m_{\phi'}$ parameter space is also ruled out below 1 MeV. We choose one BP with a  small $m_{A'}$, which satisfy the constraints.

We show the allowed regions of $m_{\phi^\prime}$-$m_{A^\prime}$ parameter space for the $\ell=e$ case in Fig.~\ref{fig:electronallowed}.}

\begin{figure}[htb]													\begin{tikzpicture}																\node (img1)  {\includegraphics[height=8cm,width=14.2cm]{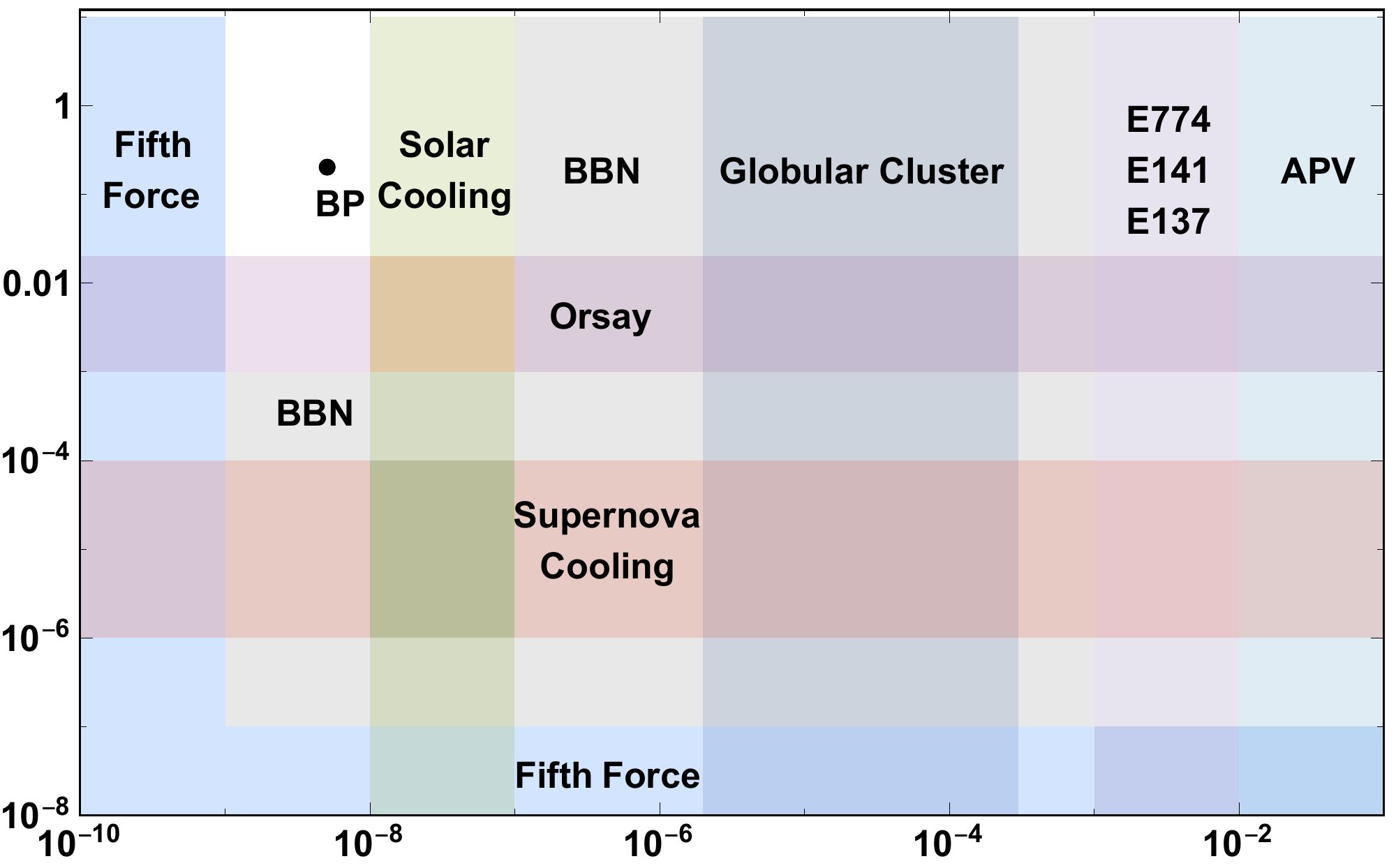}};	\node[below=of img1, node distance=0cm, yshift=0.7cm,font=\color{black}] {\textbf{\large{$ m_{A^\prime}$} (GeV)}};								 \node[left=of img1, node distance=0cm, rotate=90, anchor=center,yshift=-0.6cm,font=\color{black}] {\textbf{\large{$m_{\phi^\prime}$} (GeV)}};															\end{tikzpicture}								\caption{\label{fig:electronallowed}  Allowed regions of $m_{\phi^\prime}$-$m_{A^\prime}$ parameter space for  the case $\ell = e$ are shown as white region.   We show one BP : $m_{\phi^\prime}=$ 200 MeV and $m_{A^\prime}=$ 5 eV.  }				\end{figure}

\item{ We also can have a scenario with second-generation right-handed quarks and  second or first generation right-handed leptons. The allowed parameter space for this scenario will be similar to the previous two scenarios. In this scenario, the decay of J$/\psi$ into $\mu \mu$ $e e$ via $A^{\prime}$ can provide constraints arising from branching ratio and lepton universality~\cite{Tanabashi:2018oca,Bifani:2018zmi}.  But our choice of  $m_{A'} \sim 100~\mev$ (with $g_{T3R} \sim 10^{-2}$) for phenomenology analysis in later sections is allowed by the constraints.}
\end{itemize}

\section{Direct Detection} \label{Direct Detection}

Direct detection experiments play crucial roles in the search for dark matter particles. Traditional direct detection experiments study the nuclear recoil spectra arising from the scattering of the dark matter particles off the nuclei. Current direct detection experiments lose their sensitivity for dark matter masses below ${\cal O}(\gev)$, when the nuclear recoil energy tends to fall below the threshold. But three current direct detection experiments can provide sensitivity to the low-mass dark matter models which we consider:
\begin{itemize}
\item{{\it CRESST-III}.  CRESST-III has a relatively low recoil energy threshold and constrains the elastic spin-independent dark matter-nucleon scattering cross section to be less than $\sigmaSI \sim 10^{-35}\cm^2$ for $m \sim 200\mev$}~\cite{Abdelhameed:2019hmk}.
\item{{\it XENON1T}.  If sub-GeV dark matter is non-relativistic, it will not deposit enough nuclear recoil energy in XENON1T to exceed the threshold.  But cosmic rays can scatter off the dark matter in the halo, producing a small population of relativistic dark matter which can deposit sufficient recoil energy in XENON1T to be detected~\cite{Bringmann:2018cvk,Dent:2019krz}. For the mass range of interest, XENON1T bounds on this scenario would require either $\sigmaSI \lesssim {\cal O}(10^{-29}-10^{-30})\cm^2$ or $\sigmaSI \gtrsim {\cal O}(10^{-28})\cm^2$.}
\item{{\it CDEX-1B}. For the dark matter mass range 50-180 MeV, the Migdal effect provides the best bound on the spin-independent dark matter-nucleon scattering cross section \cite{Liu:2019kzq}}. It requires the cross section to be less than $\sigmaSI \sim 10^{-32}-10^{-34}\cm^2$ for the above mass range.
\end{itemize}

In this section, we will also study the nuclear recoil spectra for future direct detection experiments where the threshold can go down to 0.1 eV. Another way to detect the sub-GeV dark matter particle is to study the dark matter scattering off an electron. Currently, experiments like XENON10~\cite{Essig:2012yx}, SuperCDMS~\cite{Agnese:2018col} and SENSEI~\cite{Abramoff:2019dfb} can put constraints on models of low mass dark matter which scatters off electrons, but our  model parameter space is allowed by these constraints.

Our model can have both dark matter-nucleus and dark matter-electron scattering. First, we study the nuclear recoil spectra in detail and then we show electron scattering results. The relevant part of the Lagrangian for direct detection in terms of the physical fields is
\begin{dmath} \label{lagrangian}												\mathcal{L}_{int} = -\frac{1}{2}\frac{m_1}{V} \bar{\eta}_1  {\eta}_1 \phi^\prime-\frac{1}{2}\frac{m_2}{V} \bar{\eta}_2  {\eta}_2 \phi^\prime -\frac{m_u }{V}\bar{q}^u_L q^u_R \phi^\prime -\frac{m_d  }{V}\bar{q}^d_L q^d_R \phi^\prime+\frac{1}{4\sqrt{2}}\frac{m_{A^\prime}}{V}A^\prime_\mu(\bar{\eta}_1\gamma^\mu\eta_2-\bar{\eta}_2\gamma^\mu\eta_1) -\frac{1}{2\sqrt{2}}\frac{m_{A^\prime}}{V} Q_{T_{3R}} A^\prime_\mu(\bar{q}^u_R\gamma^\mu q^u_R+\bar{q}^d_R\gamma^\mu q^d_R)  .  \end{dmath}

The dark matter candidate in our model is a Majorana fermion and it has only scalar and vector interactions. Therefore we can have Spin Independent (SI) velocity-independent dark matter-nucleus scattering processes. The scalar interaction gives SI elastic scattering and the vector interaction can produce only SI inelastic scattering. Therefore our main channel of interest will be:
\begin{itemize}
\item{Elastic SI scattering ($\bar \eta \eta \bar q_L q_R$) mediated by $\phi^\prime$ exchange.}
\item{Inelastic SI scattering ($\bar \eta_1 \gamma^\mu \eta_2 \bar q_R \gamma_\mu q_R$) mediated by $A'$ exchange.  Note that, in this case, the mixing angle doesn't enter into the matrix element.  But the mass splitting does enter in the integrals over the velocity distribution.}
\end{itemize}

We can calculate the nuclear recoil spectrum for both elastic and inelastic scattering for our model. The differential event rate per unit target mass can be expressed in terms of the differential cross section as,
\begin{dmath} 																	\frac{dR}{dE_R} = \frac{N_T \rho_\eta}{m_\eta}\int_{v_{min}}^{v_{esc}}v f(v) \left(\frac{d\sigma}{dE_R} \right)d^3v , \end{dmath}
where $N_T$ is the number of target nuclei per unit mass; $\rho_\eta \simeq $ 0.3 GeV cm$^{-3}$ is the local energy density of the incoming dark matter $\eta$; $v$ is the detector frame velocity of the incoming dark matter and $f(v)$ is the corresponding normalized velocity distribution in detector frame, and; $(d\sigma / dE_R)$ is the DM-nucleus differential scattering cross section. Here $v_{min}$ is minimum dark matter velocity required for a scatter to produce recoil energy $E_R$, and $v_{esc} = 540$ km s$^{-1}$ is the local galactic escape velocity of the dark matter.

In general, the differential cross section for a dark matter particle $\eta$ of mass $m_\eta$ scattering off a target nucleus of mass $m_A$ can be written as,
\begin{equation} \label{diffcrosssec}										\frac{d\sigma}{dE_R} = \frac{m_A}{2\mu^2_{\eta A} v^2} \sigma_0(\eta A \rightarrow \eta A) \frac{m_{\phi^\prime,A^\prime}^4}{(2 m_A E_R + m_{\phi^\prime,A^\prime}^2)^2} F^2(E_R) , \end{equation}
 where $E_R$ is the recoil energy of the scattered nucleus in the lab frame; $\mu_{\eta A} = \frac{m_\eta m_A}{m_\eta+m_A}$ is the reduced mass of the $\eta$-nucleus system; $F(E_R)$ is the nuclear form factor; and $\sigma_0$ is the scattering cross section at zero momentum transfer. Details on the velocity-distribution and the nuclear form factor can be found in the appendix.

For the case of elastic scattering mediated by $\phi'$, dark matter-nucleon scattering will be largely isospin-invariant.  In this case, we can express the DM-nucleus scattering cross section at zero momentum transfer in terms of the DM-nucleon spin-independent scattering cross section at zero momentum transfer ($\sigmaSI^N$):
\bea 																		\sigma_0(\eta A \rightarrow \eta A) &=& \sigmaSI^N A^2 \frac{\mu^2_{\eta A}}{\mu^2_{\eta N}} , \eea
where $\mu_{\eta N} = \frac{m_\eta m_N}{m_\eta+m_N}$ is the reduced mass of the $\eta$-nucleon system.   But for the case of inelastic scattering mediated by $A'$, scattering is exactly  isospin-violating~\cite{Chang:2010yk,Feng:2011vu,Feng:2013fyw}, since the up and down quarks have opposite charge.  In this case, one would replace $A^2$ in the above formula  with $(A-2Z)^2$.

 Using the form of $\frac{d\sigma}{dE_R} $ from Eq.~(\ref{diffcrosssec}) we can write the differential event rate as,
 \begin{equation} \label{diffrate}												\frac{dR}{dE_R} = \frac{N_T \rho_\eta m_A \sigmaSI^N A^2}{2 m_\eta \mu^2_{\eta N}}\frac{m_{\phi^\prime,A^\prime}^4}{(2 m_A E_R + m_{\phi^\prime,A^\prime}^2)^2} F^2(E_R)\int_{v_{min}}^{v_{esc}} \frac{f(v)}{v} d^3v.  \end{equation}

Let us first consider the elastic scattering  $\eta_j A \rightarrow \eta_j A$ mediated by the scalar particle $\phi^\prime$. The dark matter nucleon SI-scattering cross section at zero momentum transfer is given as,
\begin{equation}														\sigmaSI^{scalar(p,n)} = \frac{\mu_{\eta N}^2 m_\eta^2}{\pi V^4 m_{\phi'}^4} f_{p,n}^2 \end{equation}
where ~\cite{Falk:1999mq},
\begin{equation} \frac{f_{p,n}}{m_N}=\sum_{q=u,d,s}f_{T_q}^{(p,n)}\frac{f_q}{m_q}+\frac{2}{27}\left(1-\sum_{q=u,d,s}f_{T_q}^{(p,n)}\right) \sum_{q=c,b,t}\frac{f_q}{m_q}. \end{equation}
We take $f_{u,d} = m_{u,d}$, $f_{s,c,b,t}=0$. The constants $f_{T_u}^{(p)}$, $f_{T_d}^{(p)}$ and $f_{T_s}^{(p)}$ are taken to have the values 0.019, 0.041 and 0.14, respectively~\cite{Gasser:1990ce}, and the constants $f_{T_u}^{(n)}$, $f_{T_d}^{(n)}$ and $f_{T_s}^{(n)}$ are taken to have the values 0.023, 0.034 and 0.14, respectively~\cite{Gasser:1990ce}. We thus find
\bea 																\sigmaSI^{scalar(p,n)} &\sim& (4 \times 10^{-35}~\cm^2) \left(\frac{V}{10~\gev} \right)^{-4} \left(\frac{m_{\phi'}}{100~\mev} \right)^{-4} \left(\frac{\mu_{\eta N}}{100~\mev} \right)^2 \left(\frac{m_\eta}{100~\mev} \right)^2 . \eea
The kinematics of this scattering in the laboratory frame give the threshold velocity as a function of the nuclear recoil energy,
\begin{equation} \label{vminel} v_{min} = \frac{\sqrt{2m_A E_R}}{2 \mu_{\eta A}} . \end{equation}

\begin{figure}[h]														\begin{subfigure}[b]{0.48\textwidth}\includegraphics[width=0.9\linewidth,height=5cm]{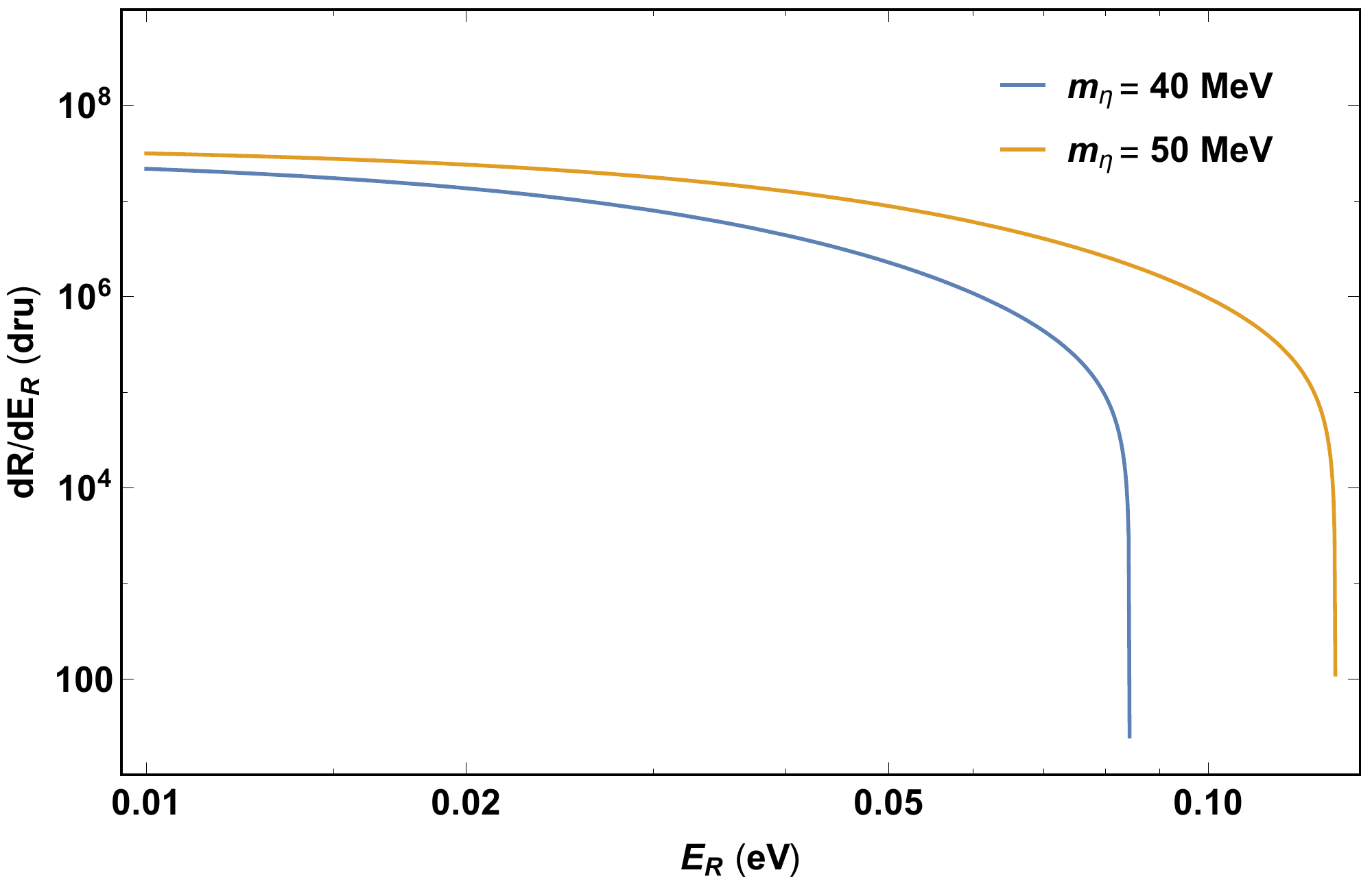}				\caption{}															\end{subfigure}															\hspace{0.2cm}														\begin{subfigure}[b]{0.48\textwidth}		\includegraphics[width=0.9\linewidth,height=5cm]{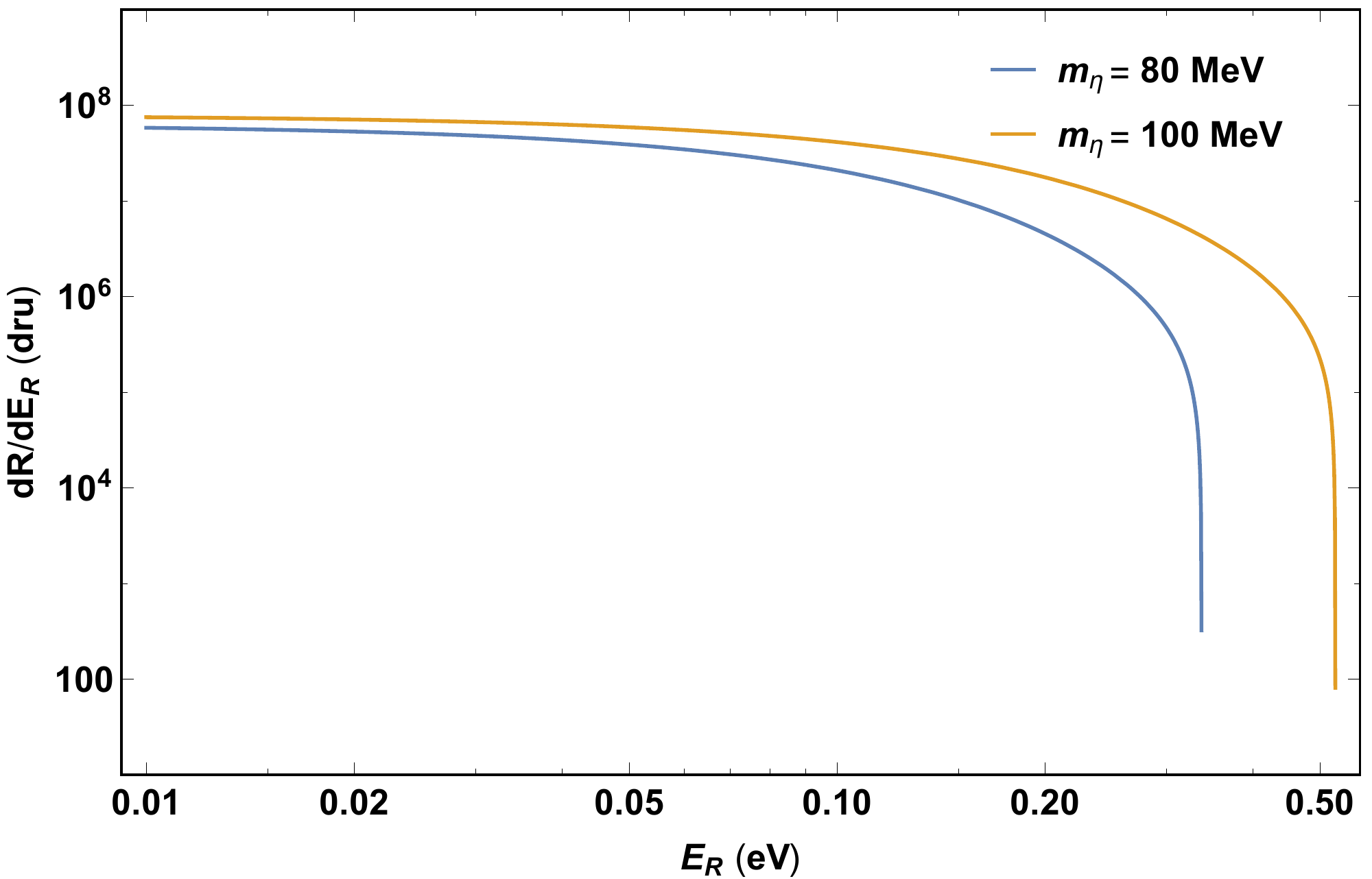}				\caption{}															\end{subfigure}												\caption{\label{fig:elastic} Differential event rate versus nuclear recoil energy for scattering off a Xenon nucleus for dark matter of various masses. The differential event rates are calculated for $m_{\phi^\prime}=$ 200 MeV and $V=$ 10 GeV. In both panels, the upper limit of recoil energy increases with the larger dark matter masses. } 															\end{figure}

We can now obtain the nuclear recoil energy spectrum for elastic scattering. We consider the elastic scattering of the $\eta$ particle off a Xenon nucleus ($A = $131 and $Z=$ 54 ). We express the differential event rate in the "differential rate unit" (dru), which is 1 event keV$^{-1}$ kg$^{-1}$ day$^{-1}$. The Fig.~\ref{fig:elastic} gives the differential event rate as a function of the nuclear recoil energy for different values of $m_\eta$. The maximum value of the nuclear recoil energy depends on the dark matter mass as we can see from Eq.~\ref{vminel}. The upper limit of $E_R$ increases with increasing dark matter mass.

\begin{figure}[h]														\begin{subfigure}[b]{0.48\textwidth}\includegraphics[width=0.9\linewidth,height=5cm]{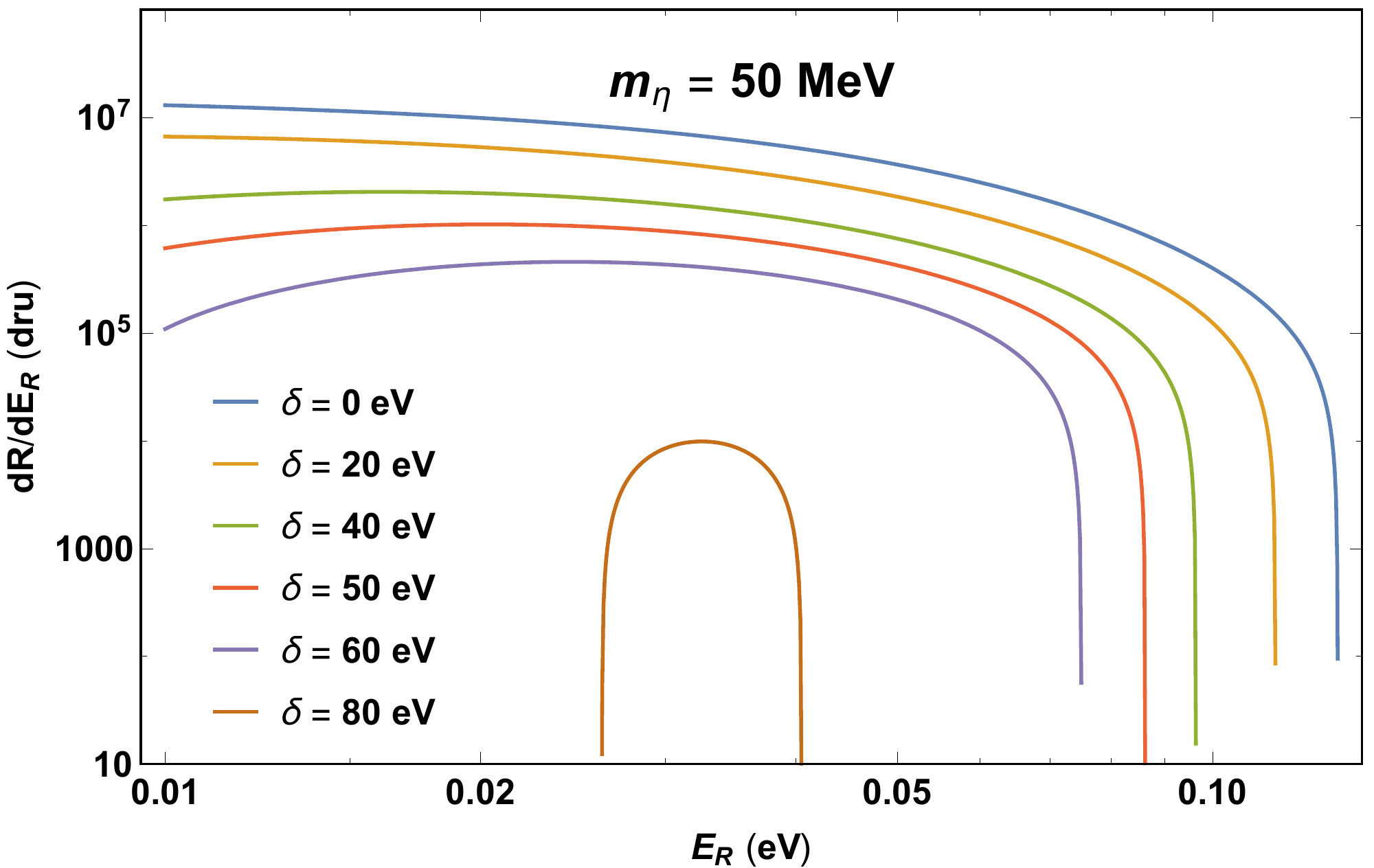}			\caption{}															\end{subfigure}															\hspace{0.2cm}														\begin{subfigure}[b]{0.48\textwidth}	\includegraphics[width=0.9\linewidth,height=5cm]{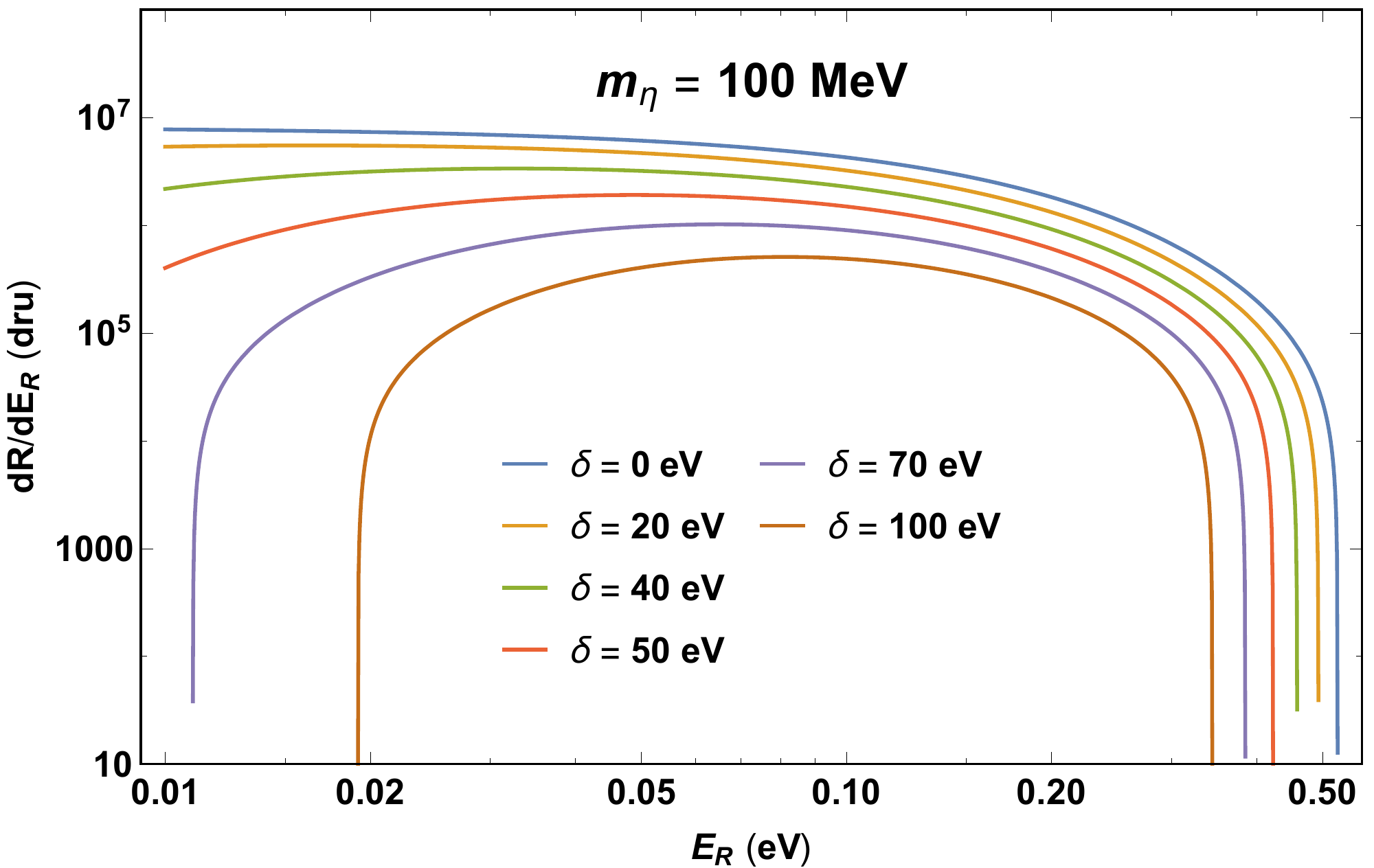}			\caption{}															\end{subfigure}												\caption{\label{fig:inelastic} Differential event rate versus nuclear recoil energy for inelastic scattering off Xenon nucleus for dark matter of various masses. The differential event rates are calculated for $m_{A^\prime}=$ 55 MeV and $V=$ 10 GeV. The values of maximum recoil energy decrease with the increasing values of $\delta$. }															\end{figure}

Let us now consider the inelastic scattering $\eta_i A \rightarrow \eta_j A$ mediated by the gauge boson $A^\prime_\mu$. We define the mass difference between two species of the dark matter particles as $\delta =$ $m_j-m_i$, and consider only the case of up-scattering ($\delta > 0$). The quantity $\delta$ enters in the kinematics of the inelastic scattering. Considering small $\delta$, we keep only the terms which are linear in $\delta$. In this limit, we can write $\mu_{\eta_j N}\simeq\mu_{\eta_i N}=\mu_{\eta N}$. For small mass splitting, the change in the matrix element is negligible, and the dominant effect is on the phase space. In particular, the threshold velocity needed in order for an inelastic scattering to yield recoil energy $E_R$ is now given by,
\begin{equation} v_{min} = \frac{1}{\sqrt{2m_A E_R}}\left( \frac{m_A E_R}{\mu_{\eta A}}+\delta \right).  \end{equation}

For a vector interaction, the zero momentum transfer dark matter-nucleon SI-scattering cross section is given by,
\bea  \sigmaSI^{vector(p, n)} &=& \frac{\mu_{\eta N}^2}{16\pi V^4} ,				\nonumber\\																		&\sim& (8 \times 10^{-36}~\cm^2) \left(\frac{V}{10~\gev} \right)^{-4} \left(\frac{\mu_{\eta N}}{100~\mev} \right)^2,  \eea in the limit of small $\delta$ in the case where dark matter couples to first generation quarks (it is one-loop suppressed otherwise).

In Fig.~\ref{fig:inelastic}, we present the recoil energy spectrum for DM-Xenon inelastic scattering for various values of $\delta$, assuming $m_\eta = 50~\mev$ (left panel) and $m_\eta = 100~\mev$ (right panel).	 The $\delta =0$ curves correspond to elastic scattering and match the shapes of the curves of the left panel and right panel of Fig.~\ref{fig:elastic} respectively. The normalization does not match because the mediator masses and couplings are different. Note, each curve terminates if $v_{min} > v_{esc}$. Larger values of $\delta$ push the nuclear recoil energy, $E_R$ to smaller values in order to satisfy the condition $v_{min} \le v_{esc} $.

\begin{figure}[h]														\begin{subfigure}[b]{0.48\textwidth}							\includegraphics[width=.9\linewidth,height=5cm]{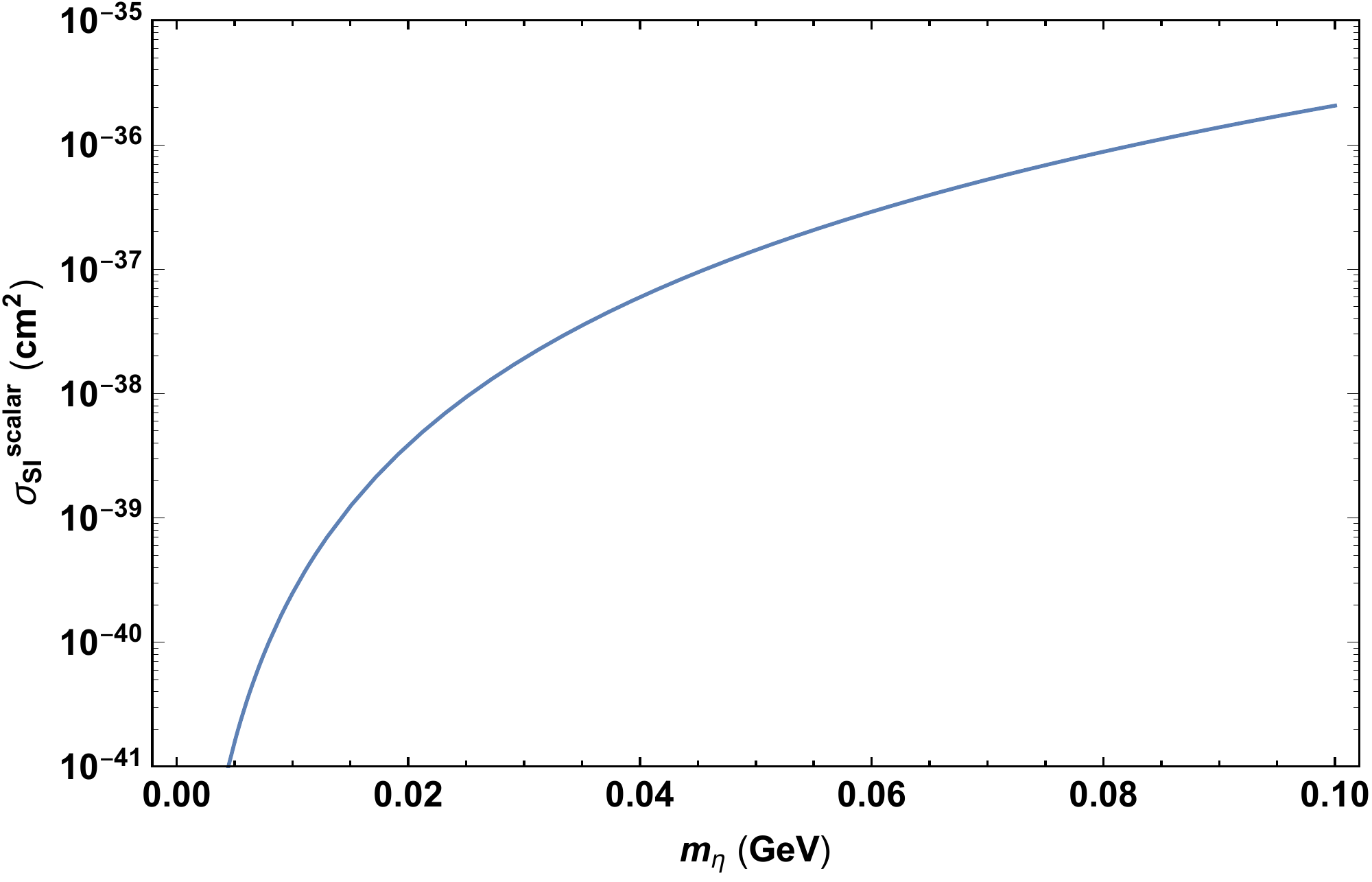}\caption{\label{fig:scalarcrosssection}}								\end{subfigure}															\hspace{0.1cm}														\begin{subfigure}[b]{0.48\textwidth}	\includegraphics[width=0.9\linewidth,height=5cm]{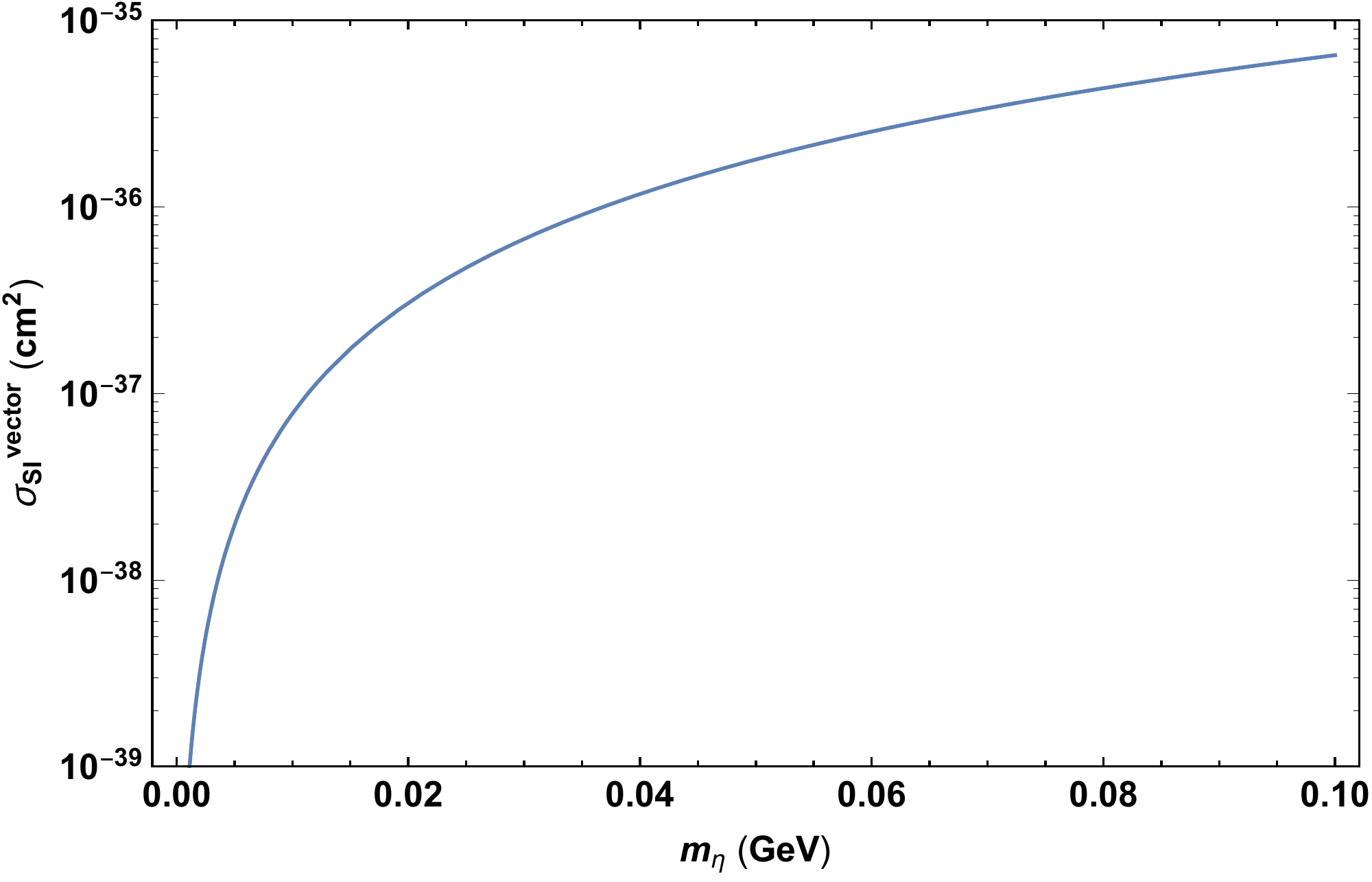}\caption{\label{fig:vectorcrosssection}}								\end{subfigure}											\caption{\label{fig:crosssection} Dark matter-nucleon scattering cross section  as a function of the dark matter mass. The cross sections are calculated for $m_{\phi^\prime}=$ 200 MeV, $\delta =0$ and $V=$ 10 GeV.	 These dark matter-nucleon cross sections are allowed by CRESST III, XENON1T and CDEX-1B constraints.}\end{figure}

\begin{figure}[h]																	\begin{subfigure}[b]{0.48\textwidth}								\includegraphics[width=.9\linewidth,height=5cm]{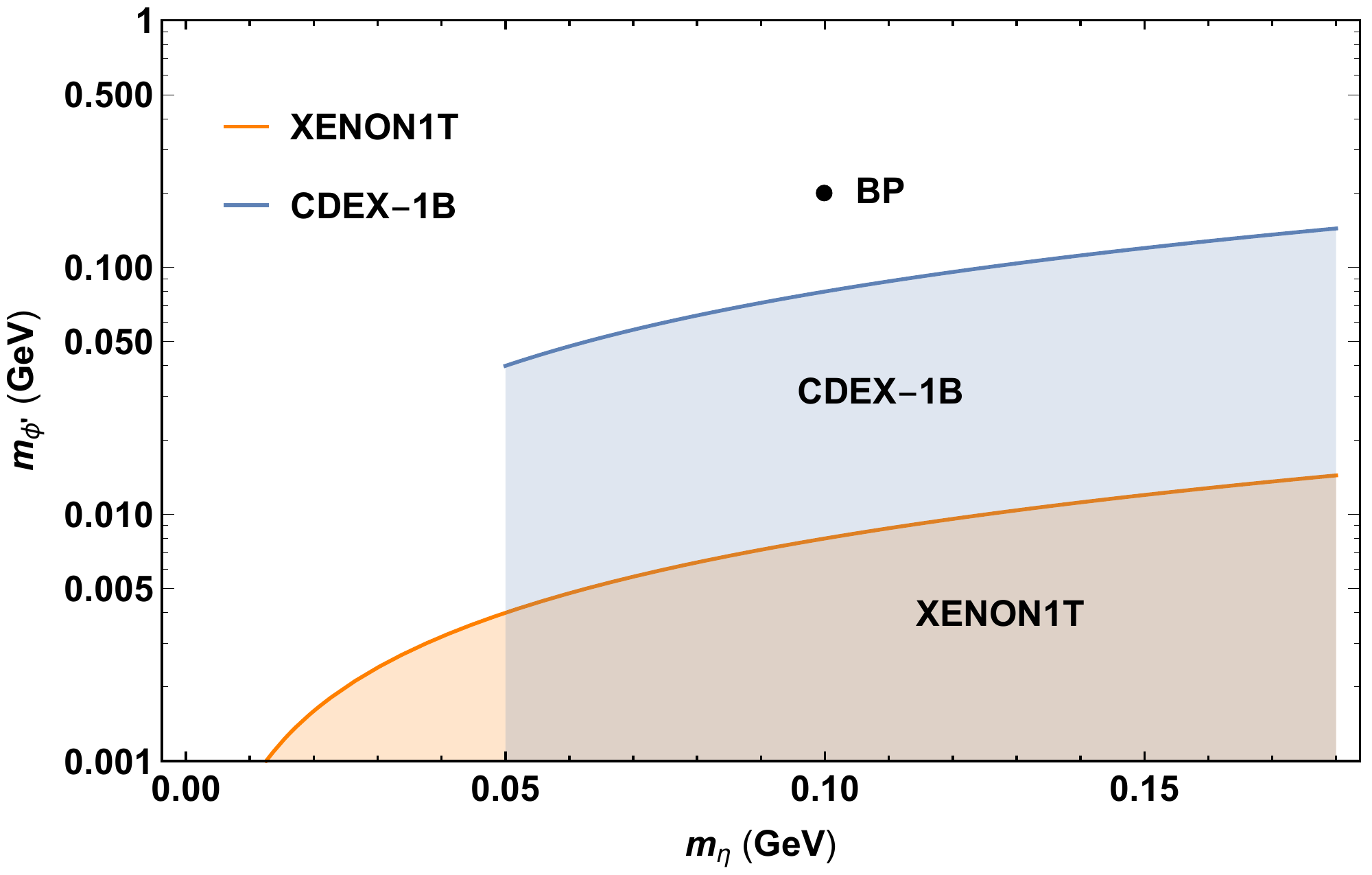}	\caption{\label{fig:Nexclusion}}										\end{subfigure}																	\hspace{0.1cm}														\begin{subfigure}[b]{0.48\textwidth}								\includegraphics[width=.9\linewidth,height=5cm]{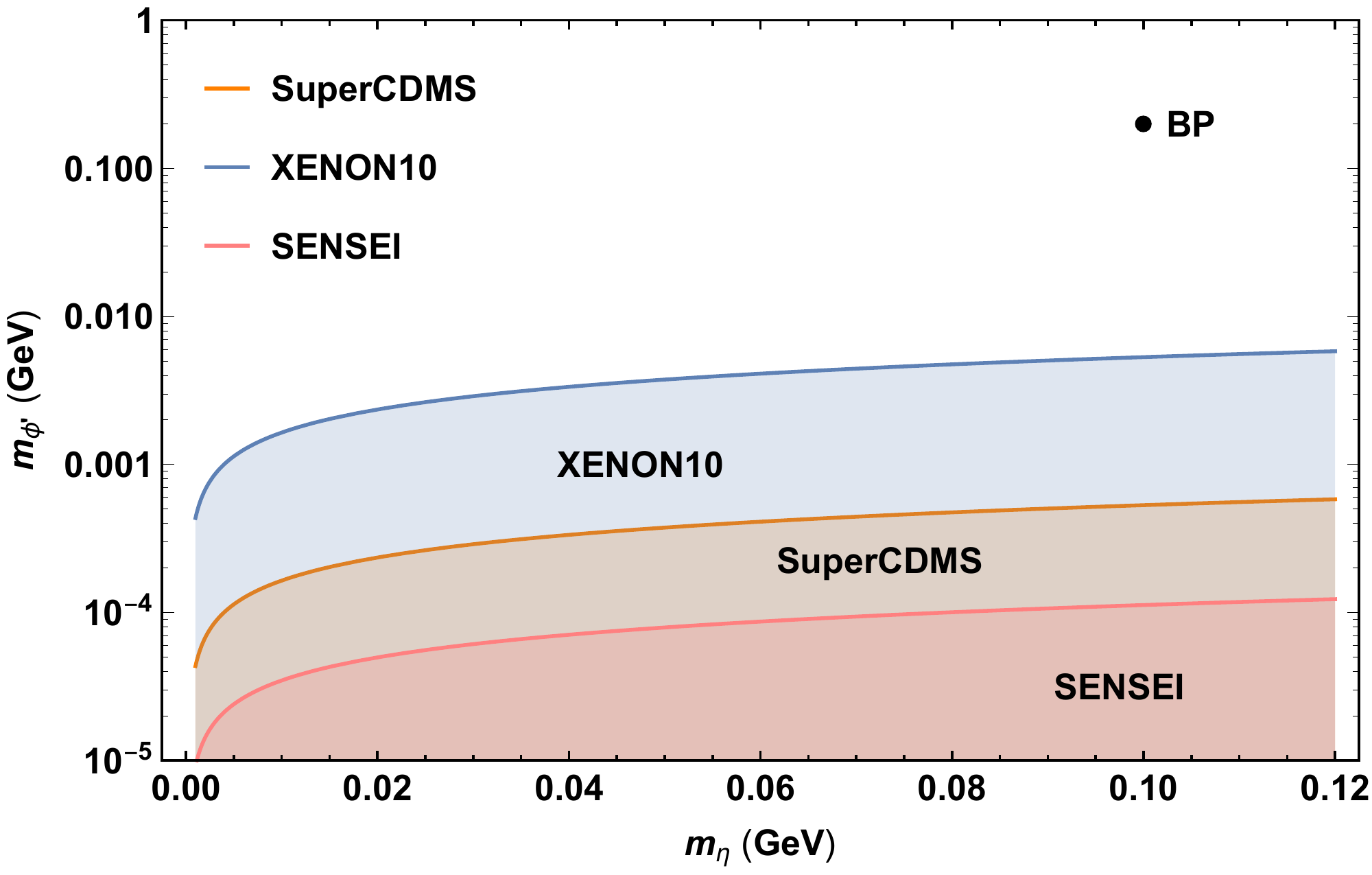}\caption{\label{fig:eexclusion}}										\end{subfigure}												\caption{\label{fig:exclusion} Exclusion plot for the $m_{\phi^\prime}$- $m_\eta$ parameter space. The left panel shows the excluded region for the dark matter-nucleon cross section bounds and the right panel shows the same for the dark matter-electron cross section constraints. In both panel, we show one BP : $m_{\phi^\prime}=$ 200 MeV, $m_\eta=$ 100 MeV and $V=$ 10 GeV.  }			\end{figure}

Fig.~\ref{fig:crosssection} shows the dark matter-nucleon scattering cross sections at zero momentum transfer for the $\phi^\prime$- and $A^\prime$- mediated processes. We have set $m_{\phi'} = 200~\mev$, and for the $A'$-mediated process, we assume $\delta=0$ (note, $\sigmaSI^{vector}$ does not depend on $m_{A'}$). Note, these scattering cross sections are well within the range allowed by XENON1T and CDEX-1B constraints. Moreover, these bounds are somewhat conservative, as the XENON1T and CDEX-1B constraints are derived assuming that the dark matter nucleon scattering cross section is equal to the zero momentum transfer cross section.  In our case, the differential scattering  cross section will be suppressed by a factor $[1+(2m_A E_R)/m_{\phi',A'}^2) ]^{-2}$. Moreover, the experimental sensitivity to $A'$-mediated scattering is suppressed by an additional factor of  $[1-(2Z/A)]^2$. For $m_\eta \leq 100~\mev$, there are no bounds from CRESST III. In Fig.~\ref{fig:Nexclusion}, we show the excluded region of the $m_{\phi^\prime}$-$m_\eta$ parameter space for the $\phi^\prime$ mediated elastic scattering corresponding to the XENON1T and CDEX-1B bounds.

\begin{figure}[htpb]													\begin{subfigure}[b]{0.48\textwidth}							\includegraphics[width=.9\linewidth,height=5cm]{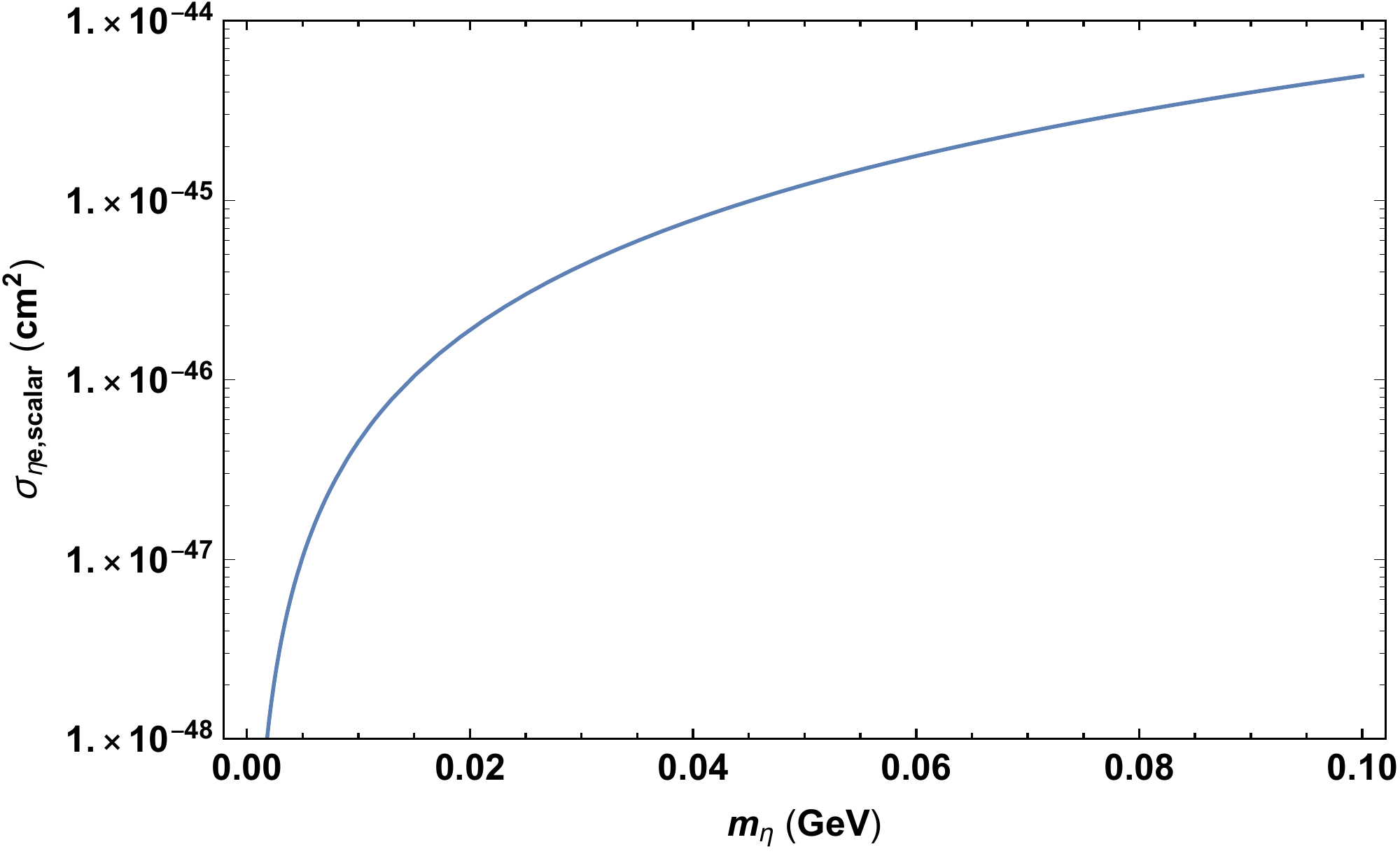}\caption{\label{fig:scalarcrosssection}}								\end{subfigure}															\hspace{0.1cm}														\begin{subfigure}[b]{0.48\textwidth}							\includegraphics[width=.92\linewidth,height=4.85cm]{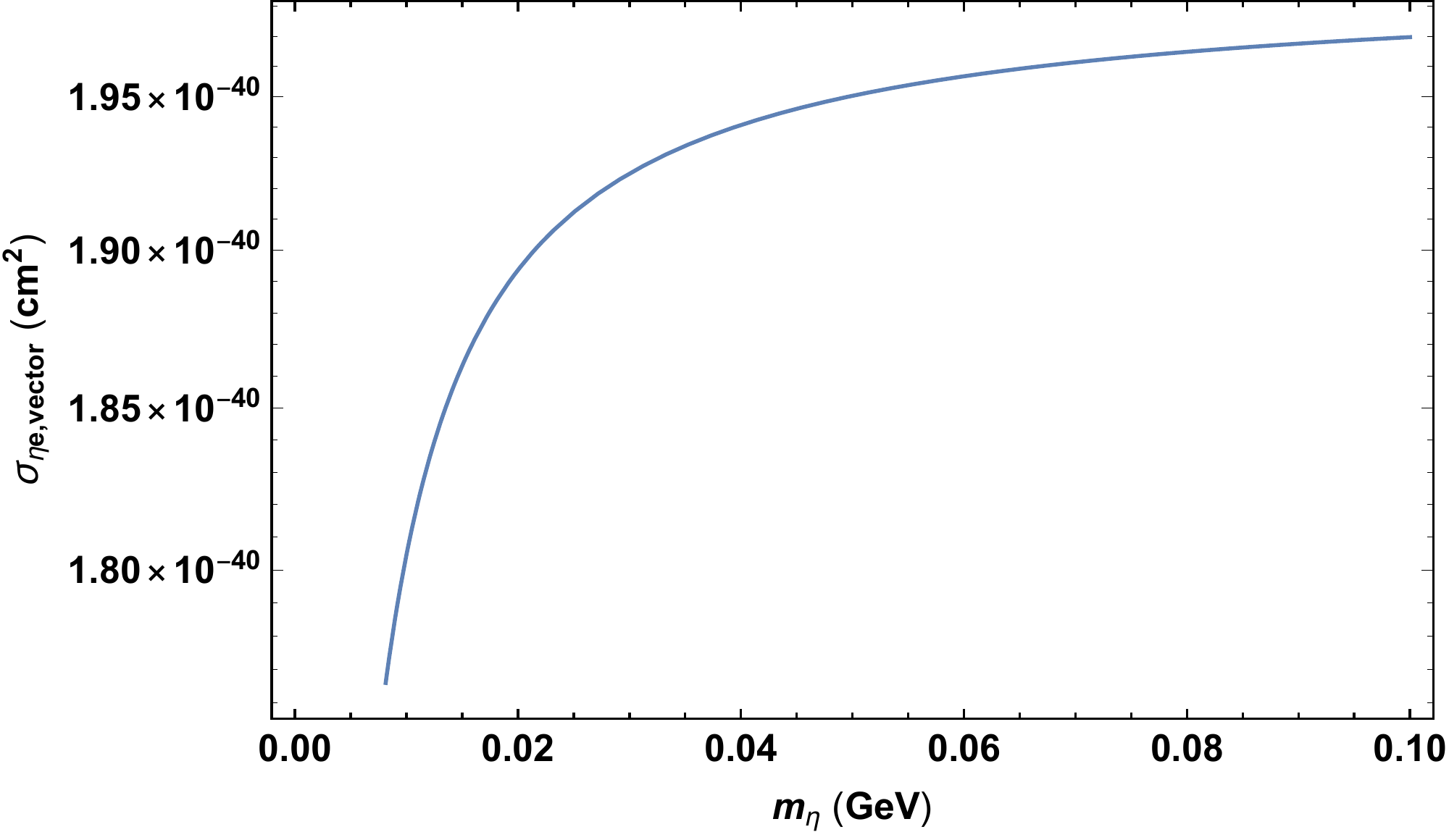}	\caption{\label{fig:vectorcrosssection}}								\end{subfigure}											\caption{\label{fig:ecrosssection} Dark matter-electron scattering cross section as a function of the dark matter mass. The cross sections are calculated for $m_{\phi^\prime}=$ 200 MeV, $\delta =0$ and $V=$ 10 GeV.}				\end{figure}

Information about dark matter and its interactions with SM particles can also be obtained from the direct detection experiments involving scattering of dark matter off electrons. For the light dark matter of mass $\mathcal {O}(1-100)$ MeV, the average energy of the incoming dark matter is $E\simeq m v^2/2 \simeq 50 $ eV$ \times (m/100 $ MeV$)$, which is sufficient for the following atomic processes:
\begin{itemize}																					\item Electron ionization (dark matter-electron scattering)									\item Electron excitation     (dark matter-electron scattering)						\end{itemize}

The typical energy required for these processes is 1-10 eV and these processes can work as visible signals in the detectors. Dark matter of mass $\mathcal {O}(1-100)$ MeV can generate these signals via scattering with the electrons. Experiments such as  XENON10~\cite{Essig:2012yx}, SuperCDMS~\cite{Agnese:2018col} and SENSEI\cite{Abramoff:2019dfb} can probe the signals generated in the dark matter-electron scattering. They put constraints on the possible scattering cross section. For the dark matter of mass  $\mathcal {O}(1-100)$ MeV, the allowed cross section is $\le 10^{-38}$ cm$^2$~\cite{Ema:2018bih, Cappiello:2019qsw}. Fig.~\ref{fig:ecrosssection} shows the dark matter-electron cross section for our model mediated via both $\phi^\prime$ and $A^\prime$.  We thus see that current experiments cannot rule out the models we are interested in here through probes of DM-electron scattering. We show the excluded region of the $m_{\phi^\prime}$-$m_\eta$ parameter space for the $\phi^\prime$ mediated elastic scattering corresponding to the XENON10, SuperCDMS and SENSEI bounds in Fig.~\ref{fig:eexclusion}.

\section{Relic density} \label{Relic density}

There are a variety of well-motivated non-standard mechanisms for obtaining the correct relic density for sub-GeV dark matter, e.g., DM production from the decay of a heavy particle~\cite{nonthermal}, freeze-in~\cite{freeze-in}, modifications to the expansion rate in the early Universe~\cite{expansion} etc. But we will focus on the more standard paradigm of a thermal relic, in which the dark matter abundance is depleted by (co-)annihilation to either Standard Model particles or to other dark sector particles.  We will assume that $m_\eta > 40~\mev$, in order to ensure that the dark matter freezes out before BBN. The dominant final states will be two-body final states, and the most relevant states are $\bar \ell \ell$, $\bar \nu \nu$, $\pi \pi$, $\pi^0 (\phi', A', \gamma)$ and the purely dark sector channels $A' A'$, $\phi' \phi'$ and $\phi' A'$.

If the mass splitting $\delta$ is large enough, then it may be that only the $\eta_1$ state is abundant at the time of freeze-out, in which case only annihilation processes are relevant for determining the relic density.  But if the mass splitting is sufficiently small, then one would expect both particles to abundant at the time of freeze-out, and co-annihilation processes will also be relevant. For co-annihilation to sufficiently deplete the dark matter abundance, the mass eigenstates $\eta_1$ and $\eta_2$ must have comparable abundances at freeze-out, implying that $\delta / m \lesssim {\cal O}(0.1)$ and that the lifetime of $\eta_2$ should be much greater than ${\cal O}(1\s)$.

For the energy range of interest to us, the tightest current constraints on dark matter annihilation arise from Planck bounds on the effect of energy injection at the time of recombination on the CMB~\cite{Aghanim:2015xee,Aghanim:2018eyx}.  If the annihilation of sub-GeV dark matter is velocity-independent and produces SM particles, then cross sections large enough to deplete the DM abundance sufficiently are generally ruled out by Planck.    To obtain the correct thermal relic density consistent with these constraints, either dark matter annihilation must either produce invisible particles  or be $p$-wave suppressed~\cite{Kumar:2013iva}. Although $p$-wave suppression only has a mild effect on the annihilation rate at the time of freeze-out, it has a dramatic effect on the annihilation rate at the time of recombination; bounds from Planck on dark matter annihilation are essentially unconstraining for the $p$-wave scenarios we consider. Alternatively, if dark matter largely co-annihilates at the time of freeze-out, but if the heavier component has decayed away by the time of recombination, then dark matter co-annihilation at the time of     recombination will be negligible, and Planck constraints will again be satisfied. If $\delta \lesssim \mev$, then the lifetime of $\eta_2$ will be much longer than the age of the Universe~\cite{Dienes:2017ylr}, but significantly shorter lifetimes are possible if $\delta > \mev$.

There is a very rich phenomenology associated with dark matter annihilation. The dominant consideration is that $A'$ couplings are suppressed by the mass of the $A'$, while $\phi'$ couplings are suppressed by the mass of the particle to which it couples. For simplicity, we will focus on two scenarios of interest:
\begin{itemize}

\item{{\it $\phi'$ resonance}: The dominant dark matter annihilation process is $\eta_i \eta_i \rightarrow \phi' \rightarrow A' A', \bar \nu \nu, \bar \ell  \ell, \pi \pi, \gamma \gamma$, where the $\phi'$ is nearly on-shell. The resonance condition is necessary to enhance the cross-section since the coupling of the $\phi'$ to the outgoing fermions is suppressed by the mass of the SM fermions.  If $A'$ is light, then its interactions are suppressed, ensuring that the dominant annihilation process proceeds through $\phi'$ production in the $s$-channel; in this case, the annihilation cross section is necessarily $p$-wave suppressed, and there are no relevant Planck bounds. In general we find
\bea																		\sigma(\eta_i \eta_i \rightarrow \phi' \rightarrow X )v_{rel} &\sim&		\frac{ m_i^2 (E^2-m_i^2)}{2 V^2 E^2[(4E^2-m_{\phi^\prime}^2)^2+(m_{\phi^\prime} \Gamma_{\phi^\prime})^2]}														\times (2m_{\phi'}  \Gamma_{\phi'}) ,										\nonumber\\																		&\sim& (9.6 \times 10^4\pb) \left( \frac{ V }{10\gev } \right)^{-2}			\left[\left( \frac{\Gamma_{\phi'}}{m_{\phi'}} \right) + \left(\frac{4E^2 - m_{\phi'}^2}{m_{\phi'}^2} \right)^2							\left( \frac{\Gamma_{\phi'}}{m_{\phi'}} \right)^{-1} \right]^{-1}			\nonumber\\																		&\,& \times														\left( \left[\frac{\langle v^2 \rangle /0.1 }{1+ \langle v^2 \rangle } \right] \frac{4m_i^2}{m_{\phi'}^2}\right) ,												\eea where $\Gamma_{\phi'}$ is the total decay width of $\phi'$.  Expressions for $\Gamma_{\phi'}$ are provided in Appendix B.

We can see that the correct relic density can only be achieved through the $\phi'$ resonance if \bea  4E^2 - m_{\phi'}^2 \gg \Gamma_{\phi'}^2 ,	\eea in which case one would need \bea	 \frac{(4E^2 - m_{\phi'}^2)^2/m_{\phi'}^4}{\Gamma_{\phi'}/m_{\phi'}} \sim 10^4 .	\eea}
\item{{\it $A'$-mediated}: If dark matter does not dominantly annihilate through a $\phi'$ mediator in the $s$-channel, then the annihilation cross section is not generally $p$-wave suppressed.  But if $m_{1,2} < m_\ell, m_\pi/2$, then no visible particles are produced at tree-level.  But if $\nu_R$ has a reasonably-sized mixing angle with a light neutrino mass eigenstate, then the dominant final state will consist of neutrinos, and Planck constraints will not be relevant.  Alternatively, if the dominant annihilation process at freeze-out is $A'$-mediated co-annihilation, and if the lifetime of $\eta_2$ is shorter than the recombination time, then Planck constraints will again be satisfied.  }
\end{itemize}

We first consider the $\ell = \mu$ scenario.  In this scenario, if $m_{\phi^\prime} \sim 50-200~\mev$, then the scalar contribution to $g_\mu-2$ is large and positive and can be tuned against the large negative contribution of the gauge boson to get the correct result.  If we accept some fine-tuning of corrections to $g_\mu-2$ from some heavy new physics, then the range $m_{\phi^\prime} \sim 10~\gev$ is also available, satisfying all other constraints.

For $m_{\phi^\prime}$ in the $50-200~\mev$ range, the dark matter abundance can be sufficiently depleted by annihilation through the $\phi^\prime$ resonance channel. In Table~\ref{tabrelic abundance}, we show one benchmark example with a dominant $\phi^\prime$ mediator channel for the muon case where  $V=10$ GeV and $m_{\nu_s}=10$ MeV.  In this case, the resonant $\phi'$ dominantly decays to $A^\prime A^\prime$, and has total decay width $\Gamma_{\phi'} = 2 \times 10^{-7}$ GeV. This scenario is not constrained by the CMB since the  $\phi^\prime$ resonance channel is $p$-wave suppressed.

For the $\ell = \mu$ scenario, we also consider the case in which $m_{\phi'} \sim 10~\gev$; processes mediated by $\phi'$ are suppressed. In Table~\ref{tabrelic abundance}, we show one example with a dominant  $A^\prime$ mediator channel for the $\ell = \mu$   case. Since we have $m_{1,2} <  m_\pi /2 $, the only processes which are available are $\eta_1 \eta_2 \rightarrow A^{\prime *} \rightarrow \bar{f} f$, where $f= \nu_s, \nu_A, e$ (if $\ell = \mu$, the $A^\prime$ will still couple to $e^+ e^-$ at one-loop through kinetic mixing). In the example, if $A^\prime$ decays to $e^+ e^-$ final state, we have constraints from CMB since the $A^\prime$ mediated channel is $s$-wave. However, we avoid that constraint if  $A^\prime$ decays to the $\nu_A \nu_A$ final state.  The branching ratio to $\nu_A\nu_A$ is larger than to the $e^+ e^-$ final state by a factor of 100. The neutrino Dirac mass is $\sim$ 10 GeV ($\lambda_{\nu_D}V=10$ GeV). The Majorana mass for $\nu_R$ is associated with a higher symmetry breaking scale (which can be lower if we introduce a Majorana neutrino mass for the left-handed neutrino). The relevant cross sections for $\eta_1 \eta_2$ annihilation  can be found in Appendix B.  Note that, unless the cross section for the process $\eta_1 \eta_2 \rightarrow e^+ e^-$ is sufficiently small, the $\eta_2$ lifetime must be significantly shorter than the recombination time in order for Planck constraints to be satisfied.

For the $\ell =e$ scenario, $m_{A^\prime}$ is constrained to lie in the sub-MeV range.  As a result, couplings to $A'$ are suppressed, and the dark matter abundance can only be depleted sufficiently if dark matter annihilates through the $\phi'$ resonance.  We show one such example in Table~\ref{tabrelic abundance}.  Since the dark matter necessarily annihilates from a $p$-wave initial state, Planck constraints are automatically satisfied.
We chose  $m_{A'}$ to be $\sim 5$ eV which makes it  not to reach equilibrium with the SM particles at early times.  But $A'$ plays no role in setting the relic density.

\begin{table}[h]																\centering																\begin{tabular}{ |c|c|c|c|c|c|c|c|c|}											\hline																			\hline																		&$m_{A'}$ (MeV)& $m_{\phi'}$ (MeV) & $m_\eta$ (MeV)&$m_{\nu_s}$(MeV)&$m_{\nu D}$(MeV)&$\langle\sigma v\rangle$ (cm$^3$/sec)&											$\sigmaSI^{scalar}$(pb)&$\sigmaSI^{vector}$(pb)\\\hline						muon case&55&200&100&10&$10^{-3}$&3$\times 10^{-26}$&2.05&6.50\\\cline{2-9}		&70&10$^4$&50&$10^{16}$&$10^4$&3$\times 10^{-26}$&3.29$\times 10^{-7}$&1.80\\ \hline																		electron case&$5\times 10^{-6}$&200&100&10&$10^{-3}$&3$\times 10^{-26}$&2.05&6.50\\\hline\hline	\end{tabular}													\caption{Masses of $A'$, $\phi^\prime$ and $\eta$ (DM) and  the corresponding thermal relic abundances are shown for muon and electron cases. The dark matter-nucleon scattering cross sections for each BP are also shown. For  the case of $A'$-mediated inelastic scattering, $\delta$ is taken small.}	\label{tabrelic abundance}\end{table}

\section{conclusion} \label{conclusion}

 The motivation of this work was to address the hierarchy problem in the light flavor sector of the SM. In order to reduce the hierarchy, we have extended the gauge symmetry of the SM with the gauge group $U(1)_{T_{3R}}$, under which  only the right-handed particles of the first two generations are charged. We have introduced a Standard Model singlet scalar field charged under $U(1)_{T_{3R}}$, which gets vev and breaks the $U(1)_{T_{3R}}$ symmetry to $Z_2$ symmetry.

 We choose the symmetry-breaking scale of $U(1)_{T3R}$ to be $\mathcal{O}(1-10)$ GeV, which allows us to obtain the $\mathcal {O} (1-100)$ MeV mass parameters for the light SM particles. We got two physical Majorana fermion $\eta_1$ and $\eta_2$, which are odd under the $Z_2$ symmetry. One or both of them can be a dark matter candidate, depending on the mass splitting between them. The mass range of the dark matter also naturally arises as $\mathcal {O} (1-100) $MeV.   For simplicity, we have chosen two specific models to work with,  one with the right-handed muon charged under $U(1)_{T3R}$ and the other with the right-handed electron charged under $U(1)_{T3R}$. Both models have first-generation right-handed quarks and a right-handed neutrino with nonzero $U(1)_{T_{3R}}$ charge (we can have a model with second-generation right-handed quarks as well). We have discussed various constraints relevant to the scale of our model.  We found an allowed region in the $m_{\phi^\prime}-m_{A^\prime}$ parameter space for both muon and electron model.

We have discussed the direct detection search for both dark matter-nucleus and dark matter-electron scattering. The dark matter fields interact diagonally with the dark scalar $\phi^\prime$ and off-diagonally with the dark gauge boson $A^\prime$. Therefore we had both elastic and inelastic SI direct detection processes. We have shown that the dark matter-nucleon cross section is allowed by  XENON1T, CDEX-1B and CRESST III constraints. The dark matter-electron scattering cross section is also allowed by the current constraints. The correct thermal relic abundance can be obtained by the standard (co-)annihilation of dark matter into the invisible SM particles or the other dark sector particles which eventually decay into SM particles. We have shown a few benchmark points, allowed by the direct detection constraints, which can give correct relic abundance and can satisfy the Plank data.

We have studied specific implementations of a general idea, which is to couple the dark sector to the light flavor sector.  In any such scenario, one would expect the energy scale of the new physics to determine the light flavor mass parameters as well as the dark matter mass, thus providing an expected mass scale for the dark matter particles -- sub-GeV.  It would be interesting to study other implementations of this idea in greater detail.

A variety of experimental efforts are being developed which have the potential to probe sub-GeV particle dark matter~\cite{Lin:2019uvt}.  Our scenario points to an interesting possibility; direct detection through inelastic scattering.  In fact, this is a generic possibility which arises in any model in which dark matter is charged under a continuous symmetry which is spontaneously broken (but under no unbroken continuous symmetries). In such a scenario, the dark particle must be a complex degree of freedom which is generically split into two real degrees of freedom with non-degenerate masses, and an interaction mediated by the massive gauge boson of the broken continuous symmetry must be inelastic.  Future direct detection experiments aimed at sub-GeV dark matter may be sensitive to inelastic scattering, but as the relevant event rates are very sensitive to the detector specifications, a detailed analysis is beyond the scope of this work, but would be an interesting future direction.

{\bf Acknowledgments}

We are grateful to Brian Batell, Tom Browder, James Dent, Ahmed Ismail, Keith Olive, Y.-Z. Qian, Sven Vahsen and Xerxes Tata for useful discussions.  We  are grateful to the organizers of the Santa Fe workshop and SUSY 2018. The work of BD and SG are supported in part by the DOE Grant No. DE-SC0010813. The work of JK is supported in part by DOE grant DE-SC0010504.

\appendix

\section{Nuclear  form factor and dark matter velocity distribution} \label{appendix1}

In this appendix, we give the nuclear form factor and the dark matter velocity distribution for the direct detection calculation.

The nuclear form factor is given by \cite{Helm:1956zz,Engel:1991wq}
\begin{equation} \label{formfactor}													F(E_R) = \frac{3 j_1(qR_1)}{qR_1}\exp \left(-q^2s^2/2 \right) ,			\end{equation}
where the momentum transferred is $q = \sqrt{2m_A E_R}$; $j_1$ is a spherical Bessel function of index 1; $s \simeq 1$  fm is the measure of nuclear skin thickness, and; $R_1 \simeq \sqrt{r^2-5s^2}$ with $r=1.2A^{1/3}$ fm and $A$ is the mass number of the target nucleus.

 We assume Maxwellian dark matter velocity distribution in the galactic rest frame \cite{Freese:1987wu}:
 \begin{equation} f(v^\prime)dv^\prime = \left[\frac{3}{2\pi v_0^2}\right]^{3/2}\exp\left(-\frac{3{v^\prime}^2}{2v_0^2}\right)4\pi{v^\prime}^2 dv^\prime , \end{equation}
 where $v_0$ has value $220$ km sec$^{-1}$. The dark matter velocity distribution is truncated at the local galactic escape velocity $v_{esc}$. To get the velocity distribution with respect to the Earth frame, we make the following Galilean transformation,
 \begin{equation}  \vec{v^\prime}=\vec{v} + \vec{v_E} ,  \end{equation}
 where $\vec{v}$ is the dark matter velocity with respect to the Earth frame and $\vec{v_E}$ is the velocity of Earth  with respect to the galactic rest frame, which is $232$ km sec$^{-1}$.  Therefore the dark matter velocity distribution in the Earth frame is given as,
 \begin{equation}  f(v)dv = \left[\frac{3}{2\pi v_0^2}\right]^{3/2} \exp\left[-\frac{3}{2v_0^2} (v^2+v_E^2)\right]\frac{v_0^2}{3vv_E}\sinh{\left(\frac{3vv_E}{v_0^2}\right)}4\pi v^2dv . \end{equation}

\section{Relic Density Details} \label{appendix2}
 In this appendix, we provide the necessary cross sections and decay widths for the relic density calculation.

\begin{itemize}

 \item {\it $\phi^\prime$-resonance}: The various partial decay widths $\Gamma_{\phi^\prime}$ are presented here:

\bea																 \Gamma_{\phi^\prime \rightarrow A^\prime_\mu A^\prime_\nu} &=& \frac{m_{\phi^\prime}^3}{64\pi V^2}\left( 1-\frac{4m_{A^\prime}^2}{m_{\phi^\prime}^2}\right)^{1/2} \left( 12\frac{m_{A^\prime}^4}{m_{\phi^\prime}^4}-4\frac{m_{A^\prime}^2}{m_{\phi^\prime}^2} +1\right),					 \nonumber\\														 \Gamma_{\phi^\prime \rightarrow f\bar{f}} &=& \frac{m_f^2 m_{\phi^\prime}}{8\pi V^2}\left(1-\frac{4m_f^2}{m_{\phi^\prime}^2} \right)^{3/2} ,				 \nonumber\\														 \Gamma_{\phi^\prime \rightarrow \bar{\eta_i}\eta_i}  &=& \frac{m_i^2 m_{\phi^\prime}}{32\pi V^2}\left(1-\frac{4m_i^2}{m_{\phi^\prime}^2} \right)^{3/2} ,															 \nonumber\\																 \Gamma_{\phi^\prime \rightarrow \gamma \gamma}  &=& \frac{\alpha^2 m_f^4}{4\pi^3V^2 m_{\phi^\prime}} \left[ 1+\left( 1-\frac{4m_f^2}{m_{\phi^\prime}^2} \right) \left( \sin^{-1}\frac{m_{\phi^\prime}}{2 m_f} \right)^2 \right]^2	\eea

\item {\it $A'$-mediated}: If $m_{1,2} < m_{A'}, m_\pi/2, m_{\phi'}$, then the only kinematically-accessible two-body final states will be $\bar f f$, where $f = \ell, \nu$.  If the dominant coupling is to $A'$, then this process can only proceed through the $s$-channel ($\eta_1 \eta_2 \rightarrow A^{'*} \rightarrow \bar f f$). We will focus on the small mass splitting limit($\delta \rightarrow 0$), in which the effect of the mass splitting is irrelevant for dark matter co-annihilation. The cross section for co-annihilation to $\bar{f} f$ final state is
\begin{equation} \sigma(\eta_i \eta_j \rightarrow A^\prime \rightarrow \bar{f} f)v_{rel} = \frac{ m_{A^\prime}^4\sqrt{E^2-m_f^2}(2E^2+m_i^2)(2E^2+m_f^2)}{96\pi V^4E^3[(4E^2-m_{A^\prime}^2)^2+(m_{A^\prime} \Gamma_{A^\prime})^2]} . \end{equation} where $E$ is the energy of the incoming dark matter particle in center-of-mass frame, and $\Gamma_{A^\prime}$ is the total decay width of the $A^\prime$ field.  The partial decay widths of $A'$ are given by,
\begin{equation}	\Gamma_{A^\prime \rightarrow f \bar{f}} = \frac{1}{24\pi V^2} \left (m_{A^\prime}^2-4m_f^2 \right)^{1/2} \left(m_{A^\prime}^2+2m_f^2 \right). \end{equation}

\end{itemize}

\bibliographystyle{plain}

\end{document}